\documentclass[aps,prb,superscriptaddress,twocolumn]{revtex4-1}

\usepackage{amsmath}
\usepackage{amsfonts}
\usepackage{amssymb}

\usepackage{bm}

\usepackage[caption=false]{subfig}
\usepackage{graphicx}
\usepackage[dvipsnames]{xcolor}

\usepackage[colorlinks=true, allcolors=blue]{hyperref}
\usepackage[capitalise]{cleveref}

\newcommand{\Var}{\mathrm{Var}}
\newcommand{\sgn}{\mathrm{sgn}}

\begin{document}

\title{Electron-pairing in the quantum Hall regime due to neutralon exchange }

\author{Giovanni A. Frigeri}
\affiliation{Max Planck Institute for Mathematics in the Sciences, D-04103, Leipzig, Germany}
\affiliation{Institut f\"{u}r Theoretische Physik, Universit\"{a}t Leipzig, D-04103, Leipzig, Germany}

\author{Bernd Rosenow}
\affiliation{Institut f\"{u}r Theoretische Physik, Universit\"{a}t Leipzig, D-04103, Leipzig, Germany}

\date{\today}

\begin{abstract}
The behavior  of electrons in condensed matter systems is mostly determined by the repulsive Coulomb interaction. However, under special circumstances the  Coulomb interaction can be effectively attractive, giving rise to electron pairing in unconventional superconductors and specifically designed mesoscopic setups. In quantum Hall systems electron interactions can play a particularly important role due to the huge degeneracy of Landau levels, leading for instance to the emergence of quasi-particles with  fractional charge and anyonic statistics. Quantum Hall Fabry-P\'{e}rot (FPI)  interferometers have attracted increasing attention due to their ability to probe such exotic physics. In addition, such interferometers are affected by electron interactions themselves in interesting ways. Recently, experimental evidence for electron pairing in a quantum Hall FPI was found (H.K. Choi et al., Nat. Comm  {\bf 6}, 7435 (2015)) . Theoretically describing an FPI in the limit of strong backscattering and under the influence of a screened Coulomb interaction, we compute electron shot noise  and indeed find a two-fold enhanced Fano factor for some parameters, indicative of electron pairing. This result is explained in terms of an electron interaction due to exchange of neutral inter-edge plasmons, so-called neutralons.

\end{abstract}


\maketitle

\section{Introduction \label{sec:Introduction}}
The quantum Hall (QH) effect is one of the most fascinating phenomena in modern condensed matter physics. It is believed that in the fractional case the elementary excitations have exotic statistics \cite{PhysRevLett.52.1583,PhysRevLett.53.722,stern2008anyons}, such that under the spatial exchange of two quasiparticles the wave function picks up a phase factor that is different from $\pm 1$ for  bosons and fermions. 
Interferometry is a promising tool for detecting such anyonic statistics in the quantum Hall regime \cite{stern2008anyons,Chamon_FPI}. 
Therefore, the investigation of quantum Hall interferometers has been an active  research field recently 
\cite{PhysRevB.72.155313,PhysRevB.76.155305,PhysRevLett.103.206806,choi2011aharonov,ABvsCD,ofek2010role,baer2013cyclic,choi2015robust,sivan2016observation,sivan2017interaction,manfra2019,Roosli2019}.
QH Fabry-P\'{e}rot interferometers (FPIs) consist of a Hall bar with two quantum point contacts (QPCs) \cite{Chamon_FPI}, which introduce a backscattering amplitude between the counter-propagating edge modes. In the limit where backscattering at the QPCs becomes strong, the FPI turns into a weakly coupled quantum dot in the QH regime. 
In addition to their potential for revealing fractional and even non-Abelian statistics \cite{nonabelian_review}, quantum Hall interferometers turned out to be an amazing tool for exploring the role of interactions \cite{ABvsCD,ofek2010role,rosenow2007influence,halperin2011theory,dinh2012influence,baer2013cyclic,choi2015robust,sivan2016observation,sivan2017interaction,Roosli2019,SciPostPhys.3.2.014,manfra2019,frigeri2019sub}.

Interestingly, indications for electron-pairing have been observed in a Fabry-P\'{e}rot interferometer (FPI) in the integer quantum Hall regime \cite{choi2015robust}. In particular, it has been observed that the Aharonov-Bohm conductance oscillations have the magnetic flux periodicity equal to half the magnetic flux quantum $h/2e$ for bulk filling factors  $2 < \nu  <5$ \cite{choi2015robust,manfra2019}, indicating that interference may be due to charge $2 e$ particles. Besides the halving of the periodicity, shot-noise measurements revealed an interfering charge equal to twice the electron charge $e^*=2e$ \cite{choi2015robust}. These observations have led the authors of Ref. \cite{choi2015robust} to suggest the formation of an electron pair in the interfering edge in order to explain the experimental findings.  On the theoretical side, the halving of the magnetic flux periodicity of the conductance has  later been explained  
by considering a model with strong edge-edge interaction and a weak bulk-edge interaction \cite{frigeri2019sub}. However, no connection between the halved flux period and electron pairing was found in that model \cite{frigeri2019sub}. Therefore, it is undoubtedly interesting to further investigate the electronic-transport in an interacting FPI in the integer quantum Hall regime and this is the main purpose of this work.
\begin{figure}
	\centering
	\includegraphics[scale=0.3]{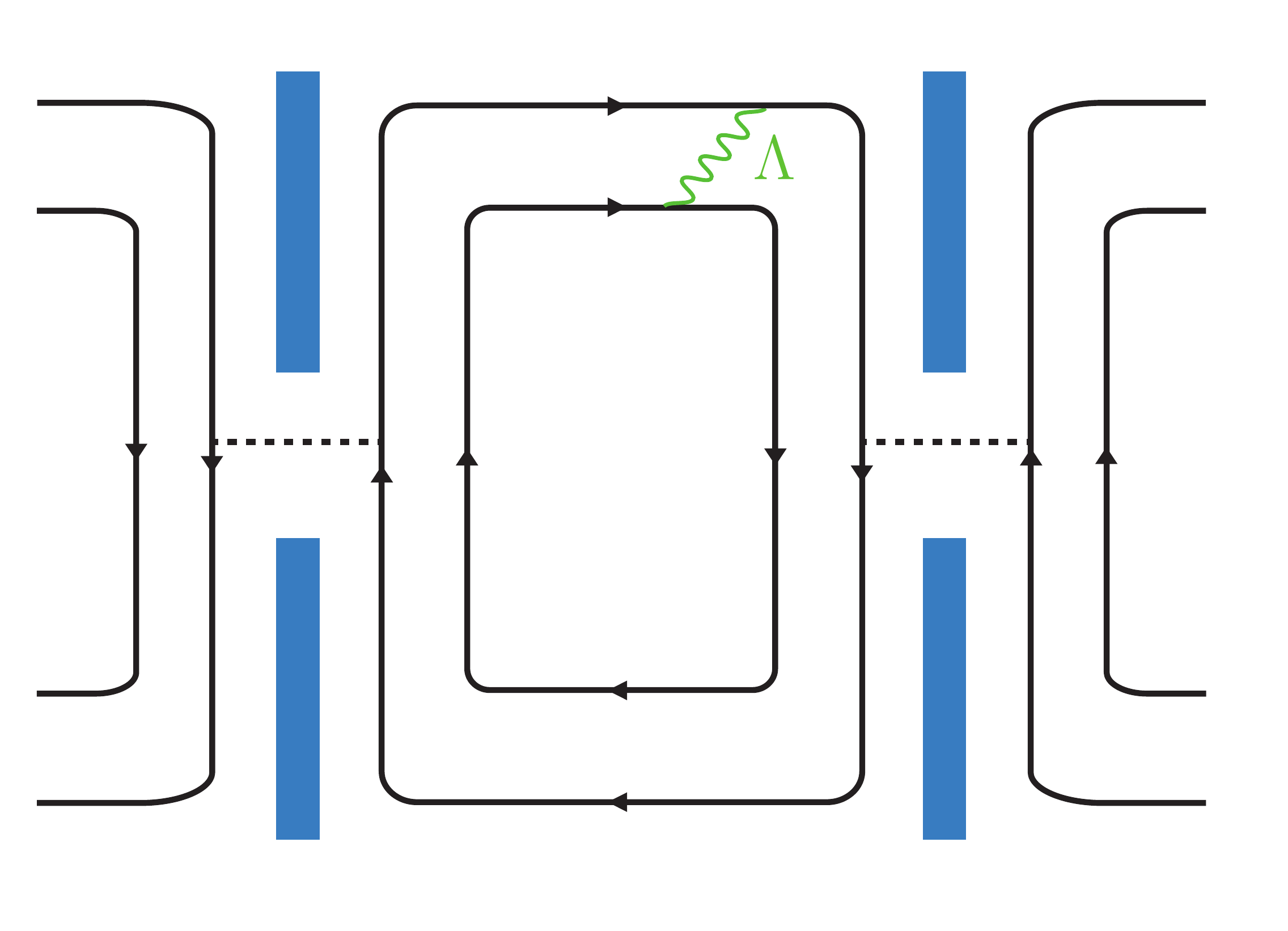}
	\caption{Schematic view of an electronic Fabry-P\'{e}rot interferometer in the closed limit, which is described by a weakly coupled quantum dot in the QH regime. Only the outermost edge interferes and the two edge modes interact in the dot region (wavy line). The parameter $\Lambda$ represents the strength of the edge-edge coupling.}\label{fig:FPI}
\end{figure}

Interactions between particles can be divided into two groups: repulsive and attractive ones. The Coulomb interaction between electrons is known to be repulsive. However,   in a variety of systems with electronic degrees of freedom it has been found that  an effective attraction between electrons arises, contrary to naive expectations \cite{Keimer+15,PhysRevLett.77.1354,Hamo,attraction_tripledot}. 
It is believed that high-temperature superconductivity can be achieved via an effective attractive interaction mediated by Coulomb repulsion \cite{Keimer+15}. However, the physics of high-temperature superconductors is rather complex,  and hence it is valuable to study the possibility of electron attraction in different and simpler systems.
For example, effective attraction between electrons due to Coulomb repulsion has been observed in quantum devices made of pristine carbon nanotubes \cite{Hamo} and in a triple quantum dot \cite{attraction_tripledot}, in which the strong repulsion between two sub-systems was exploited to make favorable the attraction between the electrons within a given sub-system.
Here, we propose a  non-equilibrium mechanism for electron-pairing mediated by repulsive Coulomb interactions, occurring in a FPI in the integer quantum Hall regime.

In this work, we consider an FPI in the presence of  inter-edge repulsive interactions and in the strong backscattering limit, i.e.~a quantum dot in the QH regime (see \cref{fig:FPI}). We choose to focus on this limit because a well-established formalism is available to calculate  shot-noise \cite{Korotkov1994,Kinaret2003,PhysRevLett.91.136801,PhysRevB.71.161301,PhysRevB.85.045325}. On the other hand, it 
is known that the limit of strong backscattering is intimately related to lowest order interference \cite{Stern+10,Roosli2019}, such that this limit provides insight into the behavior of more open interferometers as well.
 We find that the Fano factor for strongly repulsive inter-edge interactions is enhanced with respect to the Fano factor of a non-interacting interferometer. At the basis of this enhancement  is the participation of  neutral inter-edge plasmon excitations (neutralons) in  electron transport. 
 We interpret the enhancement of the Fano factor in terms of a dynamical  attraction between electrons taking place in the interfering edge via the exchange of neutralons.	 

\begin{figure}
	\centering
	\includegraphics[scale=0.37]{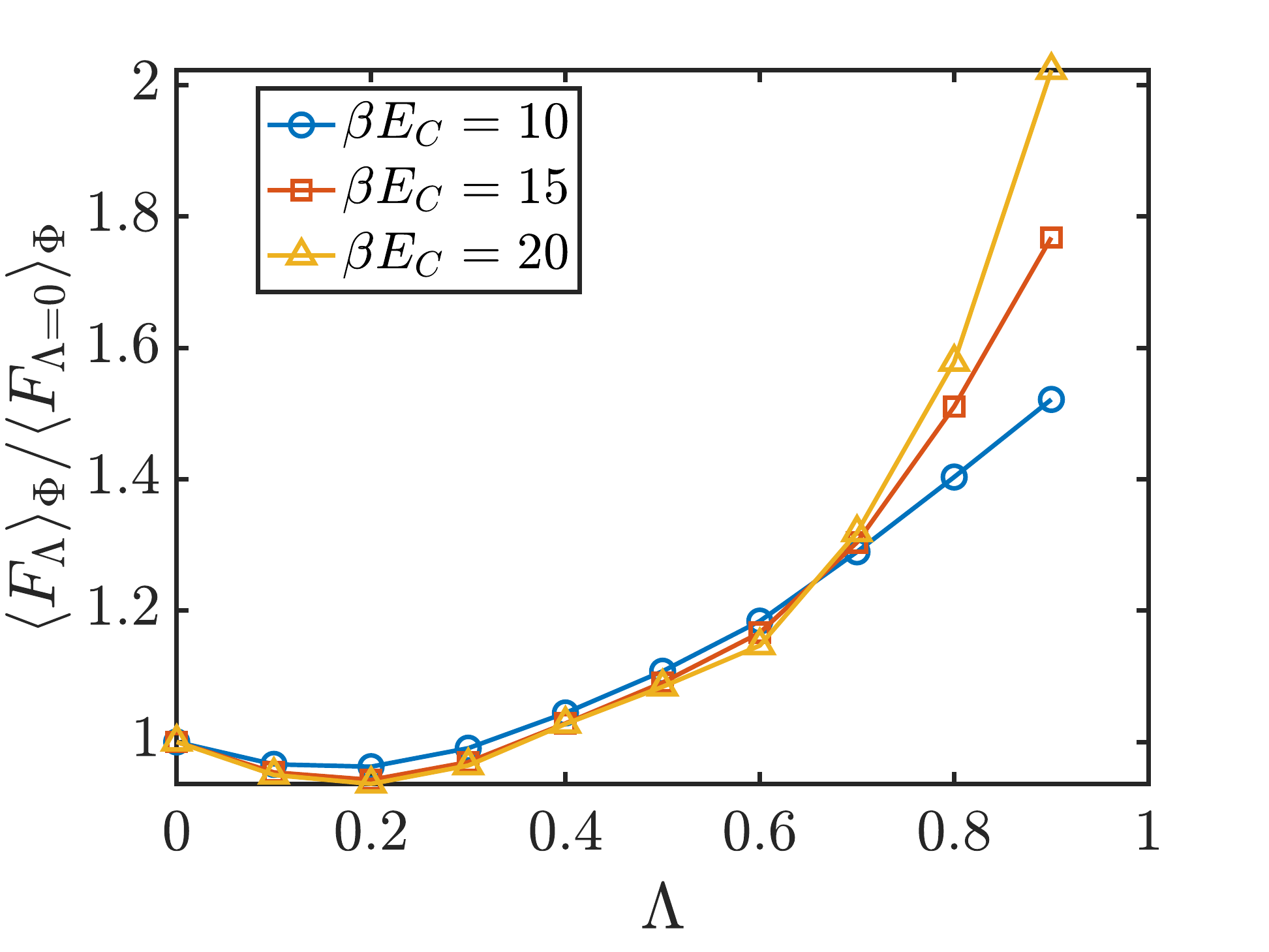}
	\caption{Ratio between the flux-averaged Fano factor $\langle F_\Lambda \rangle_\Phi$ in the presence of interaction and the non-interacting case $ \langle F_{\Lambda=0} \rangle_\Phi$ as a function of $\Lambda$ for three different temperatures ($\beta E_C = 10, 15, 20$). The bias-voltage is fixed to $eV/E_C=0.5$, where $E_c$ denotes the charging energy of the interferometer. }\label{fig:enhancement_intro}
\end{figure}

We display in \cref{fig:enhancement_intro} the magnetic flux averaged value of the Fano factor $F$  as a function of the relative  inter-edge interaction strength $0 \leq \Lambda \leq 1$, where $\Lambda = 1$ is the maximum interaction strength allowed by the requirement of electrostatic stability. 
We see that the enhancement gets stronger when increasing the coupling strength $\Lambda$, 
which can be understood as being due to the reduction $E_\sigma = (1 - \Lambda) E_C$ of the minimum neutralon excitation energy relative to the charging energy $E_C$. Hence, for large $\Lambda$, it is energetically easier to excite neutralons. Additionally, we have a bigger enhancement for lower temperatures than for higher ones, indicating that the enhancement of the Fano factor is a genuine non-equilibrium effect.  
\begin{figure}
	\centering
	\includegraphics[scale=0.37]{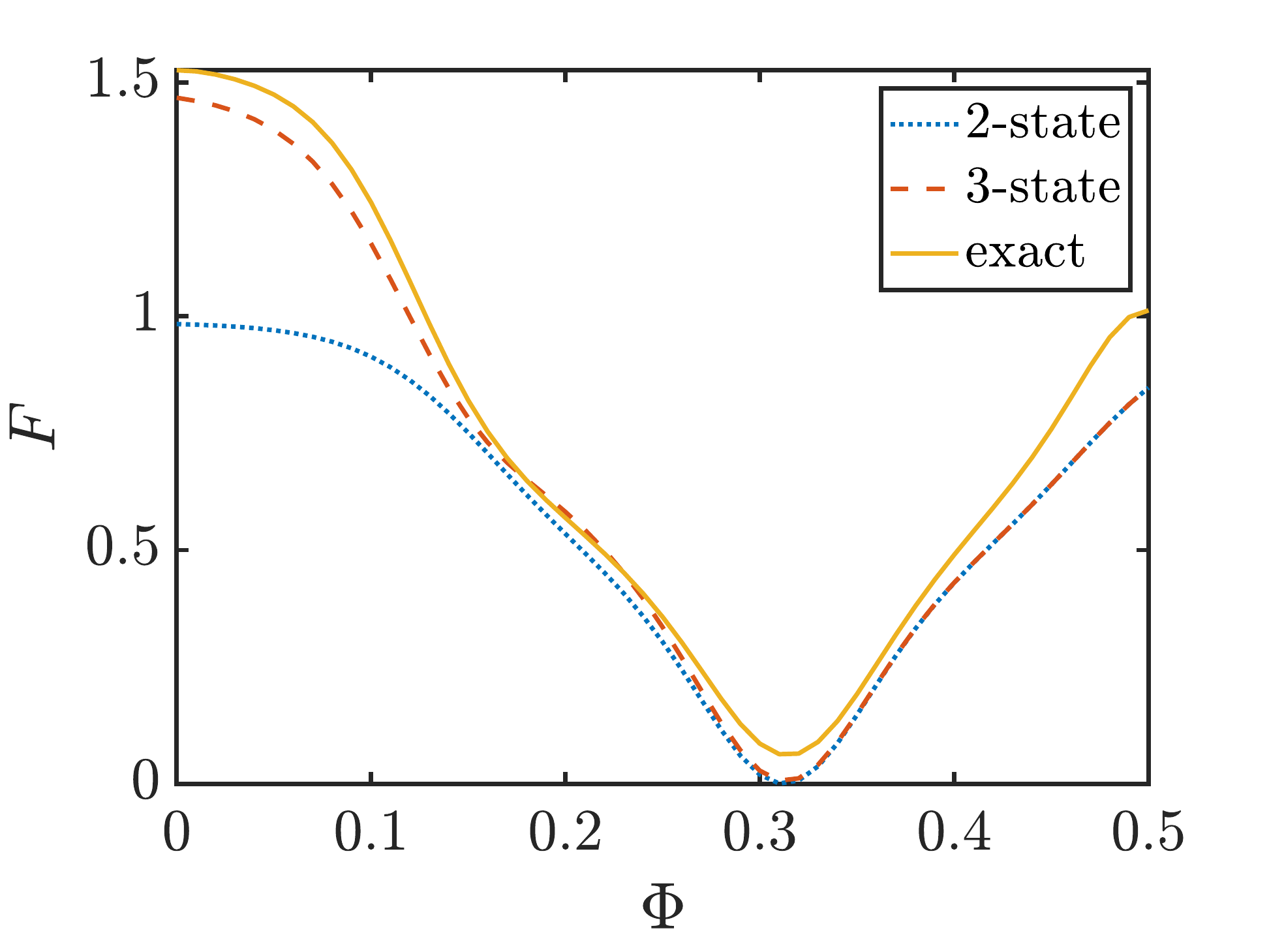}
	\caption{Fano factor as a function of the magnetic flux for a 2-state, 3-state and the exact model at $\Lambda=0.6$, $\beta E_c = 20$ and $eV/E_C = 0.5$. The Fano factor of the system is well represented by the 3-state model. }\label{fig:Fano 3 states}
\end{figure}

Electrons can move from the left lead to the right lead of the quantum dot \cref{fig:FPI} either i) independently of each other, or ii) in a correlated way in the sense that 
one electron leaves the quantum dot in an excited state, and a subsequent electron absorbs the excitation to make entering the dot easier. In case i),  the sequence between states is
\begin{equation}\label{eq:seq}
	\lvert 0, \{0\}_\sigma \rangle \rightarrow \lvert 1, \{0\}_\sigma \rangle \rightarrow \lvert 0, \{0\}_\sigma \rangle
	\rightarrow \lvert 1, \{0\}_\sigma \rangle \rightarrow \lvert 0, \{0\}_\sigma \rangle.
\end{equation}
Here, for a state $\lvert n, \{m\}_\sigma \rangle$, $n$ denotes the numer of additional electrons in the dot, and $m$ denotes the occupancy of the energetically lowest neutralon excitation.  
In case ii), a neutralon takes part in the electronic-transport. As before, one electron tunnels through the left QPC into the dot. However, a neutralon is now created in the dot when the electron exits. A subsequent electron tunnels more easily into the dot by absorbing the neutralon, and then exits the dot without leaving behind an excitation. Accordingly, we are led to consider  the sequence of transitions 
\begin{equation}\label{eq:neutr}
	\lvert 0, \{0\}_\sigma \rangle \rightarrow \lvert 1, \{0\}_\sigma \rangle \rightarrow \lvert 0, \{1\}_\sigma \rangle
	\rightarrow \lvert 1, \{0\}_\sigma \rangle \rightarrow \lvert 0, \{0\}_\sigma \rangle.
\end{equation}
\begin{figure*}
	\centering
	\includegraphics[scale=0.4]{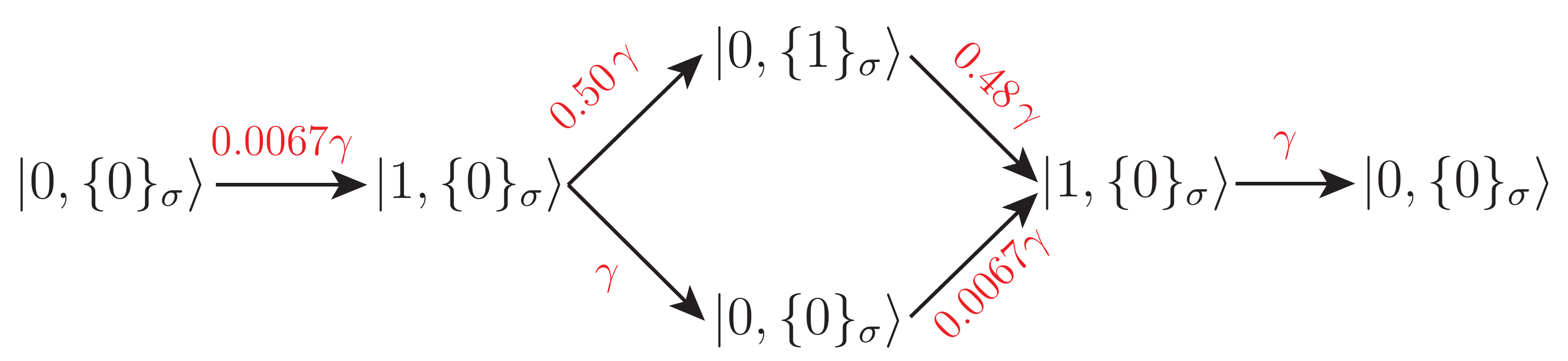}
	\caption{The two different transport channels when the neutral plasmon excitation is involved. We report the states taking part into the transport of electrons and we indicate on the arrows the respective transition rates at $\Lambda=0.6$, $\Phi=0$, $\beta E_c = 20$ and $eV/E_C = 0.5$.}\label{fig:transition rates}
\end{figure*}
In order to understand the relative contribution of processes i) and ii) to transport current and excess noise, we compute the relative probability of the two processes for a special choice of parameters  $\Lambda=0.6$, $\beta E_C = 20$, $eV/E_C = 0.5$, $\Phi = 0$. 
For these parameters, a model taking into account only the three states $\lvert 0, \{0\}_\sigma \rangle$, $\lvert 0, \{1\}_\sigma \rangle$, and
$\lvert 1,\{0\}_\sigma \rangle$ describes the Fano factor with good accuracy,  
see  \cref{fig:Fano 3 states}.  The  transition rates in processes i) and ii) are displayed in \cref{fig:transition rates}. 

We see that when the system is in  state $\lvert 1, \{0\}_\sigma\rangle$ and the electron leaves the dot, the system reaches the state $\lvert 0, \{0\}_\sigma\rangle$ with a rate that is almost twice as large as the transition rate to $\lvert 0, \{1\}_\sigma\rangle$. 
However, subsequently the rate for a second electron to enter the dot is almost two orders of magnitude larger for the state $\lvert 0, \{1\}_\sigma\rangle$ as compared to the  state $\lvert 0, \{0\}_\sigma\rangle$ without a neutralon. This big difference is due to the fact that the neutralon provides the necessary energy for a lead electron to enter the dot level in the first case, whereas  only few electrons in the tail of the Fermi distribution have an energy sufficiently large for entering the dot in the second case. As a consequence, in the case with a neutralon excitation present in the dot, the second electron tunnels through the dot almost instantaneously, such that the tunneling of two electrons takes approximately the same time as the tunneling of a single  electron in the case without a neutralon excitation. 

When neglecting correlated tunneling of three and more electrons (justified by extra factors of $1/2$ for each additional electron tunneling in a  correlated fashion), we can adopt an effective description in which either single electrons can tunnel with probability $p_{\rm single}$ or pairs of electrons with probability $ p_{\rm pair}$. Due to the ratio of transition rates discussed above, we know that 
$p_{\rm single} = 2 p_{\rm pair}$, and taking into account the sum rule $p_{\rm single} + p_{\rm pair} = 1$ we find the values $p_{\rm single} = 2/3$ and 
$p_{\rm pair} = 1/3$. 

When evaluating the Fano factor for a stochastic process describing tunneling of single electrons and pairs (for details see \cref{sec:Electron attraction mediated by plasmon excitations}), we find that the Fano factor is an increasing function of $p_{\rm pair}$. In the limit where  only single electrons tunneling is possible ($p_{\rm pair}=0$), we obtain the Fano factor for a Poisson process, $F=1$. On the other hand, we obtain $F=2$ in the opposite limit when $p_{\rm pair}=1$ because only electron pairs tunnel now and we have a Poisson process but with charge carrier $e^*=2e$. The Fano factor takes intermediate values, $1 < F < 2$, when tunneling of single electrons and electron pairs coexists. In particular, for the parameters discussed above, we obtain from the stochastic model described in section \ref{sec:Electron attraction mediated by plasmon excitations} $F=1.5$ (red star in \cref{fig:Fano pair}). In conclusion, this example gives insight into the mechanism responsible for the enhancement of the Fano factor: a neutralon excitation left behind by one electron makes it easier for a second electron to enter the dot, and thus gives rise to correlated tunneling, comparable to pairing of electrons due to a retarded interaction. 
We emphasize that for even larger values of $\Lambda > 0.6$, correlated processes involving larger number of electrons and  neutralons need to be taken into account and can lead to a Fano factor in excess of two even when $p_{\rm pair} < 1$.

The reminder of the paper is organized as follows: in \cref{sec:Interacting Fabry-Perot interferometer} we introduce the model used to describe the closed limit of an FPI with repulsive inter-edge coupling. In \cref{sec:Formalism}, we set up the master equation to understand the effect of interactions on the properties of the FPI. The results for the conductance, the noise and the Fano factor in the presence of strong edge-edge interaction are discussed in \cref{sec:Enhancement of the Fano factor from repulsive interaction}. We compare in \cref{sec:Electron attraction mediated by plasmon excitations} our results with a model in which the tunneling of single electrons and pairs is possible. Finally, we discuss in \cref{sec:Coherent tunneling} how  the results for an interferometer in the closed limit obtained via the master equation are related to the case of more open interferometers described by a scattering formalism, and comment on the difficulty of extracting an effective charge for interferometers in the closed limit. 
\section{Model \label{sec:Interacting Fabry-Perot interferometer}}
In the following we describe a model which allows to study   how shot  noise is influenced by the presence of repulsive Coulomb interactions.
We consider an FPI in the closed limit, consisting of two edge modes (filling factor $\nu=2$) coupled by a local repulsive interaction among each other. 
The interferometer is composed of left ($L$) and right ($R$) semi-infinite leads and a finite-size central dot ($D$) region. Electron tunneling from the leads to the dot, and vice versa, occurs only into  the outermost edge mode. 
We assume that the two edges in the dot region enclose the same  magnetic flux $\Phi = AB/\phi_0$, where $A$ is the area of the interferometer, $B$ the perpendicular magnetic field and $\phi_0=h/e$ the magnetic flux quantum. We are interested in the case of a strongly  screened bulk-edge interaction \cite{choi2015robust,manfra2019,frigeri2019sub}, which allows us to neglect the bulk of the system in the following.
The setup is depicted in~\cref{fig:FPI}, and we focus on the consequences of an  edge-edge interaction  on transport properties. 
The $L/R$ lead is at equilibrium with the respective reservoir, described by temperature $T$ and voltage $V_{L/R}$. We characterize the $L/R$ reservoir with the respective Fermi function
\begin{equation}\label{eq:Fermi function}
	f_{L/R}(E) = \frac{1}{1+\exp\left[(E-eV_{L/R})/k_B T\right]}\ .
\end{equation}
A voltage bias $V=V_L-V_R$ is applied between the reservoirs, driving the system out of equilibrium.
%

The left and right leads are semi-infinite, and since an inter-edge interaction in the lead region would only renormalize tunneling matrix elements, we neglect it here.  Then, the Hamiltonian for the $l = L,R$ lead is ($\hbar=1$)
\begin{equation}\label{eq:lead hamiltonian}
	H_l = \frac{v}{4\pi} \int_{0}^{\infty} dx \left[\left(\partial_x\phi_{1,l}\right)^2 + \left(\partial_x\phi_{2,l}\right)^2\right],
\end{equation}
where we assumed that the edge modes have the same velocity $v$ and $\phi_{1/2,l}$ are the  bosonic fields
describing edge modes in the  $l$ lead which are coupled/un-coupled to the dot region. The fields $\phi_{j,l}$ ($j=1,2$) satisfy   commutation relations
\begin{equation}\label{eq:commutation relation phi}
	\left[\phi_{j,l}(x), \phi_{j,l}(x^\prime)\right] = i \pi\,\sgn\left(x-x^\prime\right),
\end{equation}
and are related to  one-dimensional electron densities via
\begin{equation}\label{eq:one-dimensional density}
	\rho_{j,l}(x) = \frac{\partial_x \phi_{j,l}(x)}{2\pi} \ .
\end{equation}
\\
On the other hand, a local  repulsive interaction couples the edge modes inside the dot region. The appropriate Hamiltonian $H_D$ describing the finite-size dot  is
\begin{align}
	H_D = \frac{v}{4\pi} \int_{0}^{L} dx \big[&\left(\partial_x\phi_{1,D}\right)^2 + \left(\partial_x\phi_{2,D}\right)^2  \nonumber\\
		&+2\Lambda \left(\partial_x\phi_{1,D}\right) \left(\partial_x\phi_{2,D}\right)\big], \label{eq:H dot}
\end{align}
where $L$ is the size of the dot, $\phi_{1/2,D}$ are the bosonic field associated with the outer and inner edge mode in the dot, and $\Lambda$ describes the relative strength of the edge-edge coupling ($0 \leq \Lambda \leq 1$). 
The fields $\phi_{1/2,D}$ satisfy the commutation relation  \cref{eq:commutation relation phi}, and are related to the electron density via \cref{eq:one-dimensional density}.
In order to diagonalize the Hamiltonian \cref{eq:H dot}, we perform a  change of basis 
\begin{equation}\label{eq:rotation}
	\begin{pmatrix}
	\phi_\rho \\
	\phi_\sigma
	\end{pmatrix} = \frac{1}{\sqrt{2}} 
	\begin{pmatrix}
	1 & 1 \\
	-1 & 1
	\end{pmatrix}
	\begin{pmatrix}
	\phi_{1,D} \\
	\phi_{2,D}
	\end{pmatrix},
\end{equation}
allowing us to rewrite \cref{eq:H dot} in  diagonal form as 
\begin{equation}
	H_D = \frac{v}{4\pi} \int_{0}^{L} dx \left[\left(1+\Lambda\right) \left(\partial_x \phi_\rho\right)^2 + \left(1-\Lambda\right) \left(\partial_x \phi_\sigma\right)^2\right].
\end{equation}
The fields $\phi_{\rho/\sigma}$ are decomposed into a part $\phi^p_{\rho/\sigma}$  obeying periodic boundary conditions, and a non-periodic part zero mode part  $\phi^0_{\rho/\sigma}$, which is needed to obtain the correct commutation relations  \cref{eq:commutation relation phi} in a finite size system \cite{Geller1997}
\begin{equation}\label{eq:decomposition operator phi}
	\phi_{\rho/\sigma}(x) = \phi_{\rho/\sigma}^p(x) + \phi_{\rho/\sigma}^0(x) \ .
\end{equation}
The periodic part $\phi_{\rho/\sigma}^p$ can be expanded in terms of bosonic operators $b_{k,\rho/\sigma}$, $b^\dagger_{k,\rho/\sigma}$, that annihilate or create  a charge/neutral plasmon with momentum $k = 2\pi m/L$ ($m=1,2,\dots$)
\begin{equation}\label{eq:phi p}
	\phi_{\rho/\sigma}^p(x) = \sum_{k>0} \sqrt{\frac{2\pi}{kL}} e^{-k \alpha/2} \left(b_{k,\rho/\sigma} e^{ i k x}+h.c\right),
\end{equation}
where $\alpha$ is a short distance cut-off. The zero mode part $\phi_{\rho/\sigma}^0$ can be expressed as
\begin{equation}\label{eq:phi 0}
	\phi_{\rho/\sigma}^0(x) = \frac{2\pi}{L}N_{\rho/\sigma} x-\chi_{\rho,\sigma} \ ,
\end{equation}
where $N_{\rho/\sigma}$ is the number operator for the charge/neutral sector respectively, and $\chi_{\rho,\sigma}$ is an Hermitian operator canonically conjugate to $N_{\rho/\sigma}$, such that $[\chi_{\rho,\sigma}, N_{\rho/\sigma}] = i$. 
In this way, the Hamiltonian in \cref{eq:H dot} takes the form
\begin{equation}
	H_D = \sum_{\eta=\rho,\sigma} \left(\frac{E_\eta}{2} N^2_\eta + \sum_{m=1}^{+\infty} m\, E_\eta\, b_{m,\eta}^\dagger b_{m,\eta}\right),
\end{equation}
where we used the momentum quantization $k = 2\pi m/L$, and we defined the energies
\begin{equation}\label{eq:epsilon rho/sigma}
	E_{\rho/\sigma} = E_C \left(1 \pm \Lambda\right) \ .
\end{equation}
Here, the charging energy $E_C$ is related to the edge velocity and the size of the dot via $E_C=2\pi v/L$.
We obtain the final Hamiltonian for the dot by expressing $N_{\rho/\sigma}$ in terms of the number operator of the outer interfering edge $N_1$ and the inner non-interfering edge $N_2$ with the transformation in \cref{eq:rotation}, and including the magnetic flux $\Phi$ via the substitution $N_{1/2} \to N_{1/2}-\Phi$ \cite{Geller1997,rosenow2007influence,halperin2011theory}
\begin{align}\label{eq:dot Hamiltonian}
	H_D = \frac{E_\rho}{4} \left(N_1+N_2-2\Phi\right)^2 + \frac{E_\sigma}{4} \left(N_1-N_2\right)^2 \nonumber\\
	+ \sum_{m=1}^{+\infty} m \left[E_\rho b_{m,\rho}^\dagger b_{m,\rho} + E_\sigma b_{m,\sigma}^\dagger b_{m,\sigma}\right].
\end{align}
Because of the presence of QPCs at positions $x_L$ and $x_R$, an electron in the outermost edge can tunnel from/to the leads 
onto and off the dot.  The tunneling Hamiltonian is
\begin{equation}\label{eq:tunneling hamiltonian}
H_T = \sum_{l = L,R} \left[\tilde{t}_l \psi^\dagger_{1,l}(x_l) \psi_{1,D}(x_l) + h.c\right],
\end{equation}
such that an electron is annihilated in the outermost edge of the dot, and created in the outermost edge of the $l$ lead, with amplitude $\tilde{t}_l$. The hermitian conjugate term describes the inverse process (from to lead to the dot). The total Hamiltonian of the interacting Fabry-P\'{e}rot interferometer is then
\begin{equation}\label{eq:hamiltonian FPI filling 2}
H = H_L+H_R+H_D+H_T \ ,
\end{equation}
with the individual terms given by \cref{eq:lead hamiltonian,eq:dot Hamiltonian,eq:tunneling hamiltonian}.
\section{Rate Equation Formalism \label{sec:Formalism}}
In this section, we introduce the formalism with which we calculate the observables relevant for describing the transport properties of an FPI in the closed limit.
\subsection{Tunnelling rates \label{subsec:Tunnelling rates}}
By inspection of the Hamiltonian   \cref{eq:dot Hamiltonian} it is clear that we can describe  the state of the dot by the quantum numbers
\begin{equation}
	\lvert N_1, N_2, \{n\}_\rho, \{n\}_\sigma \rangle,
\end{equation}
where $N_1$ is the number of electrons on the outermost edge, $N_2$ the number of electrons on the inner edge, and $\{n\}_\rho = \{n_{1,\rho}, n_{2,\rho}, \dots\}$, $\{n\}_\sigma = \{n_{1,\sigma}, n_{2,\sigma}, \dots\}$ are  the occupation numbers of the charge and neutral plasmon modes, respectively. Accordingly, the energy of the dot state $\lvert N_1, N_2, \{n\}_\rho, \{n\}_\sigma \rangle$ is
\begin{align}\label{eq:energy filling2}
	E\left(N_1, N_2, \{n\}_\rho, \{n\}_\sigma\right) = \frac{E_\rho}{4} \left(N_1+N_2-2\Phi\right)^2 \nonumber\\
	+ \frac{E_\sigma}{4} \left(N_1-N_2\right)^2 + \sum_{m=1}^{+\infty} m \left[E_\rho n_{m,\rho} + E_\sigma n_{m,\sigma} \right],
\end{align}
with $E_{\rho/\sigma}$ defined in \cref{eq:epsilon rho/sigma}. 
Since we want to be in a regime where the electron number on the outer edge is a good quantum number, the tunneling amplitudes $|\tilde{t}_l|$ need to chosen in such a way that the tunneling conductance into or out of the outer edge is much smaller than the conductance quantum $e^2/h$. The rate for the transition from an initial state $| i \rangle$ to a final state $| f \rangle$ due to the small perturbation $H_T$ can be calculated using  Fermi's golden rule \cite{nazarov2009quantum}
\begin{equation}\label{eq:golden rule}
\Gamma\left(| i \rangle \rightarrow | f \rangle\right) = 2\pi \left|\langle i \left|H_T\right|f\rangle\right|^2 \delta\left(E_f-E_i\right).
\end{equation}
By using \cref{eq:tunneling hamiltonian} with \cref{eq:golden rule}, and taking into account that the electrons in the leads have a thermal distribution,  we find that the total tunneling rates for adding or removing an electron to or from the outermost edge mode through the $l=L,R$ QPC are
\begin{align}
&\Gamma_l \left(N_1, N_2, \{n\}_\rho, \{n\}_\sigma \rightarrow N_1+1, N_2, \{n^\prime\}_\rho, \{n^\prime\}_\sigma\right) \nonumber\\
&= \gamma_l M(\{n\}_\rho, \{n^\prime\}_\rho) M\left(\{n\}_\sigma, \{n^\prime\}_\sigma\right) \nonumber\\
&\qquad\times f_l\left(\Delta_+ \left(N_1, N_2, \{n\}_\rho, \{n\}_\sigma, \{n^\prime\}_\rho, \{n^\prime\}_\sigma\right)\right), \label{eq:addition rate filling2}\\
\nonumber\\
&\Gamma_l\left(N_1, N_2, \{n\}_\rho, \{n\}_\sigma \rightarrow N_1-1, N_2, \{n^\prime\}_\rho, \{n^\prime\}_\sigma\right) \nonumber\\
&= \gamma_l M(\{n\}_\rho, \{n^\prime\}_\rho) M\left(\{n\}_\sigma, \{n^\prime\}_\sigma\right) \nonumber\\
&\qquad \times \left[1-f_l\left(-\Delta_- \left(N_1, N_2, \{n\}_\rho, \{n\}_\sigma, \{n^\prime\}_\rho, \{n^\prime\}_\sigma\right)\right)\right],\label{eq:removing rate filling2}
\end{align}
where we defined the bare tunneling rate $\gamma_l \equiv 2\pi |\tilde{t}_l|^2 \rho_F/L$,    with density of states $\rho_F= 1/2 \pi \hbar v$, and $f_l$ denoting the Fermi function of the $l$ lead in \cref{eq:Fermi function}.
The matrix element $M\left(\{n\}, \{n^\prime \}\right)$ accounting for the possible  excitation of plasmons  is
\begin{align}\label{eq:matrix element filling 2}
&M(\{n\}, \{n^\prime\}) \nonumber\\
&= \frac{\displaystyle \prod_{m=1}^{m_{\rm max}} e^{-1/2m} \left(\frac{1}{2m}\right)^{|n_m-n_m^\prime|} \frac{n_m^{(<)}!}{n_m^{(>)}!} \left[L_{n_m^{(<)}}^{|n_m-n_m^\prime|}\left(\frac{1}{2m}\right)\right]^2}{\displaystyle\prod_{m=1}^{m_{\rm max}} e^{-1/2m}}.
\end{align}
where $n_m^{(>)} = \max(n_m, n_m^\prime)$, $n_m^{(<)} = \min(n_m, n_m^\prime)$, $L_a^b(x)$ are the associated Laguerre polynomials \cite{abramowitz1972handbook} and $m_{\rm max}$ is an index such that $n_m = n_m^\prime= 0$ for every $m>m_{\rm max}$. We want to remark that generally the value of $m_{\rm max}$ in \cref{eq:matrix element filling 2} is different for the excitations in the $\sigma$ and $\rho$ sectors.
A detailed derivation of \cref{eq:matrix element filling 2} can be found in \cref{app:suppl calculation of matrix element} \cite{Kinaret2003,kim2006nonequilibrium}.
The addition and subtractions energies are  obtained from \cref{eq:energy filling2} as
\begin{align}
&\Delta_\pm \left(N_1, N_2, \{n\}_\rho, \{n\}_\sigma, \{n^\prime\}_\rho, \{n^\prime\}_\sigma\right) \nonumber\\
&= E\left(N_1\pm 1, N_2, \{n^\prime\}_\rho, \{n^\prime\}_\sigma\right) - 
E\left(N_1, N_2, \{n\}_\rho, \{n\}_\sigma\right).
\end{align}
\subsection{Master equation \label{subsec:Master equation nu=2}}
Knowledge of the transition rates for electron tunneling allows us to set up a rate equation to describe the transport properties of the interferometer, and understand the effect of interactions. The time evolution of the probability that the system is in configuration $N_1, N_2, \{n\}_\rho, \{n\}_\sigma$ at time $t$ is governed by the master equation \cite{Furusaki1998,Kinaret2003,kim2006nonequilibrium}
\begin{widetext}
	\begin{align}
	&\frac{d}{dt} p\left(N_1, N_2, \{n\}_\rho, \{n\}_\sigma, t\right) \nonumber\\
	&= \sum_{l=L,R} \sum_{\{n^\prime\}_\rho, \{n^\prime\}_\sigma} \bigg\{
	\Gamma_l \left(N_1+1, N_2, \{n^\prime\}_\rho, \{n^\prime\}_\sigma \rightarrow N_1, N_2, \{n\}_\rho, \{n\}_\sigma \right) 
	p\left(N_1+1, N_2, \{n^\prime\}_\rho,  \{n^\prime\}_\sigma, t \right)\nonumber\\
	&\hspace{3cm} +\Gamma_l \left(N_1-1, N_2, \{n^\prime\}_\rho, \{n^\prime\}_\sigma \rightarrow N_1, N_2, \{n\}_\rho, \{n\}_\sigma \right) 
	p\left(N_1-1, N_2, \{n^\prime\}_\rho,  \{n^\prime\}_\sigma, t \right)\nonumber\\
	&\hspace{3cm} -[\Gamma_l \left(N_1, N_2, \{n\}_\rho, \{n\}_\sigma \rightarrow N_1+1, N_2, \{n^\prime\}_\rho, \{n^\prime\}_\sigma \right)\nonumber\\
	&\hspace{3.5cm}+ \Gamma_l \left(N_1, N_2, \{n\}_\rho, \{n\}_\sigma \rightarrow N_1-1, N_2, \{n^\prime\}_\rho, \{n^\prime\}_\sigma \right)] 
	p\left(N_1, N_2, \{n\}_\rho, \{n\}_\sigma, t \right) \bigg\},\label{eq:master equation}
	\end{align}
\end{widetext}
with the transition rates $\Gamma_l$ given by \cref{eq:addition rate filling2,eq:removing rate filling2}. 
In the long time limit, the system reaches a stationary state with probability distribution 
\begin{equation}
\lim\limits_{t\to\infty} p\left(N_1, N_2, \{n\}_\rho, \{n\}_\sigma, t \right) = p_{st}\left(N_1, N_2, \{n\}_\rho, \{n\}_\sigma \right).
\end{equation}
The stationary probability distribution satisfy the condition
\begin{equation}\label{eq:normalization condition}
\sum_{N_1}\sum_{\{n\}_\rho, \{n\}_\sigma} p_{st}\left(N_1, N_2, \{n\}_\rho, \{n\}_\sigma \right) = p_{st}\left(N_2\right).
\end{equation}
However, we cannot compute $p_{st}\left(N_2\right)$ within our model, because we assume that the tunneling occurs only in the outermost edge mode. In principle, we would need to specify the dynamics of innermost edge mode and include the matrix elements for changing the value of $N_2$ to solve the master equation in \cref{eq:master equation}. In our model the innermost edge mode $N_2$ is a slow variable and it has no influence on the pairing mechanism here described. In addition, it is connected to a thermal bath and therefore we approximate the stationary probability distribution of $N_2$ as
\begin{equation}\label{eq:p N2}
	p_{st}\left(N_2\right) = p_{eq}\left(N_2\right),
\end{equation}
with the thermal distribution
\begin{equation}
p_{eq}\left(N_2 \right) = \frac{\sum_{N_1, \{n\}_\rho \{n\}_\sigma} e^{-\beta E(N_1, N_2, \{n\}_\rho, \{n\}_\sigma)}}{\sum_{N_1, N_2, \{n\}_\rho \{n\}_\sigma} e^{-\beta E(N_1, N_2, \{n\}_\rho, \{n\}_\sigma)}},
\end{equation}
with the energy given by \cref{eq:energy filling2} and inverse temperature $\beta = 1/k_BT$. Taking into account the assumption in \cref{eq:p N2}, it is now possible to solve the master equation in \cref{eq:master equation}.

\subsection{Matrix formulation of the master equation}
It order to compute the transport current and  noise in an efficient way, it is  useful to recast the previous equations in a matrix notation \cite{kim2006nonequilibrium}. To that end, we collect the state probabilities $p\left(N_1, N_2, \{n \}_\rho, \{n \}_\sigma, t\right)$ into the probability vector $\bm{p}(t)$ 
\begin{equation}\label{eq:probability vector filling2}
\left(\bm{p}(t)\right)_{\{ N_1, N_2, \{n\}_\rho, \{n\}_\sigma\}} = p\left(N_1, N_2, \{n\}_\rho, \{n\}_\sigma, t\right),
\end{equation}
and we define the matrices $\bm{\Gamma}_l^{(\pm)}$ and $\bm{\Gamma}_l^{(0)}$
\begin{align}\label{eq:addition matrix filling2}
&\left(\bm{\Gamma}_l^{(\pm)}\right)_{\{N_1\pm 1, N_2, \{n^\prime\}_\rho, \{n^\prime\}_\sigma\}; \{N_1, N_2, \{n\}_\rho, \{n\}_\sigma\}} \nonumber\\
&= \Gamma_l\left( N_1, N_2, \{n\}_\rho, \{n\}_\sigma  \rightarrow  N_1\pm 1, N_2, \{n^\prime\}_\rho, \{n^\prime\}_\sigma\right), \\
&\left(\bm{\Gamma}_l^{(0)}\right)_{\{N_1, N_2, \{n\}_\rho, \{n\}_\sigma\}; \{N_1, N_2, \{n\}_\rho, \{n\}_\sigma\}} \nonumber\\
&=\sum_{\{n^\prime\}_\rho, \{n^\prime\}_\sigma} \Bigg[ \left(\bm{\Gamma}_l^{(+)}\right)_{\{N_1+1, N_2, \{n^\prime\}_\rho, \{n^\prime\}_\sigma\}; \{N_1, N_2, \{n\}_\rho, \{n\}_\sigma\}}  \nonumber\\
& + \left(\bm{\Gamma}_l^{(-)}\right)_{\{N_1-1, N_2, \{n^\prime\}_\rho, \{n^\prime\}_\sigma \}; \{N_1, N_2, \{n\}_\rho, \{n\}_\sigma\}} \Bigg],
\label{eq:loss matrix filling2}
\end{align}
with the transition rates given by \cref{eq:addition rate filling2,eq:removing rate filling2}. The column index of the matrices $\bm{\Gamma}_l^{(\pm)}$ represents the initial configuration, while the row index represents the final state. The matrix $\boldsymbol{\Gamma}_l^{(0)}$ is a diagonal matrix, whose $n$-th element is obtained by summing all the elements in the $n$-th column of the $\boldsymbol{\Gamma}_l^{(\pm)}$ matrices.\\
In this way, the master equation in \cref{eq:master equation} can be cast in the compact form
\begin{equation}\label{eq:master equation matrix}
\frac{d}{dt} \boldsymbol{p}(t) = -\boldsymbol{\Gamma}\boldsymbol{p}(t),
\end{equation}
with the transition matrix $\boldsymbol{\Gamma}$, generating the temporal evolution, defined as
\begin{equation}\label{eq:Gamma matrix}
\boldsymbol{\Gamma} \equiv \sum_{l = L,R} \left(\boldsymbol{\Gamma}_l^{(0)}-\boldsymbol{\Gamma}_l^{(+)}-\boldsymbol{\Gamma}_l^{(-)}\right),
\end{equation}
and we require \cref{eq:p N2} to hold.
The solution to \cref{eq:master equation matrix} is
\begin{equation}\label{eq:solution master equation}
\bm{p}(t) = e^{-\bm{\Gamma}(t-t_0)} \bm{p}(t_0),
\end{equation}
and the stationary probability distribution is given by
\begin{equation}\label{eq:stationary solution}
\bm{p}_{st} = \lim\limits_{t\to\infty} \bm{p}(t).
\end{equation}
\subsection{Average current, noise and Fano factor}
The average current through the left/right QPC at time $t$ is given by 
\begin{equation}
\langle I_{L/R} (t) \rangle =  e \sum_{\substack{N_1, N_2 \\\{n\}_\rho, \{n\}_\sigma}} \left[\left( \bm{\Gamma}^{(+)}_{L/R} -\bm{\Gamma}^{(-)}_{L/R}\right) \bm{p}(t)\right]_{\{N_1, N_2, \{n\}_\rho, \{n\}_\sigma\}},
\end{equation}
and the total average current flowing trough the system is $\langle I(t) \rangle = (\langle I_L(t) \rangle - \langle I_R(t) \rangle)/2$. As a consequence of current conservation, the total stationary current $\langle I \rangle$ is related to the stationary current $\langle I_{L/R} \rangle$ through the left/right QPC, $\langle I \rangle  = \langle I_L \rangle = -\langle I_R \rangle$. Accordingly, we can obtain the stationary current $\langle I \rangle$ as
\begin{equation}\label{eq:stationary current ME}
\langle I \rangle =  e \sum_{\substack{N_1, N_2 \\\{n\}_\rho, \{n\}_\sigma}} \left[\left( \bm{\Gamma}^{(+)}_{L} -\bm{\Gamma}^{(-)}_{L}\right) \bm{p}_{st}\right]_{\{N_1, N_2, \{n\}_\rho, \{n\}_\sigma\}}.
\end{equation}

In addition to the average current, we also need an expression for the fluctuations of the current in time in order to obtain the Fano factor.  To this end, we define the noise 
\begin{equation}\label{eq:noise}
S = \lim_{t\to\infty} 2 \int_{-\infty}^{+\infty} d\tau \left[\langle I(t+\tau) I(t)\rangle - \langle I \rangle^2\right].
\end{equation}
The noise $S$ can be computed from the correlation function $K_{l m}(\tau)$ between the left and right currents 
\begin{equation}
K_{lm}(\tau) = \lim_{t\to\infty}\langle I_l(t+\tau) I_m(t)\rangle - \langle I_l \rangle \langle I_m \rangle,
\end{equation}
which can be expressed for $\tau \geq 0$ as \cite{Korotkov1994}
\begin{widetext}
	\begin{align}
	K_{l m}(\tau) = -\langle I_l \rangle \langle I_m \rangle + e^2 \sum_{\substack{N_1, N_2 \\\{n\}_\rho, \{n\}_\sigma}} \Bigg\{\bigg[  &\left( \bm{\Gamma}^{(+)}_{l}-\bm{\Gamma}^{(-)}_{l} \right) e^{-\bm{\Gamma}\tau} \left( \bm{\Gamma}^{(+)}_{m}-\bm{\Gamma}^{(-)}_{m} \right) +\delta_{lm} \delta(\tau) 
	\left( \bm{\Gamma}^{(+)}_{l}+\bm{\Gamma}^{(-)}_{l} \right) \bigg]\bm{p}_{st}\Bigg\}_{\{N_1, N_2, \{n\}_\rho, \{n\}_\sigma\}}, \label{eq:noise ME matrix}
	\end{align}
\end{widetext}
and the case $\tau \leq 0$ can be obtained by using $K_{lm}(\tau) = K_{ml}(-\tau)$.
Finally, the Fano factor $F$ is defined as the ratio between the excess noise $S-S_{eq}$  and the average current $\langle I \rangle$ obtained from \cref{eq:stationary current ME}
\begin{equation}\label{eq:Fano factor ME}
F = \frac{S-S_{eq}}{2 e \langle I \rangle},
\end{equation}
where $S_{eq} = 4k_BT G$ is the Nyquist-Johnson noise, evaluated at zero voltage and temperature $T$, related to the conductance $G$ via the fluctuation-dissipation theorem \cite{nazarov2009quantum}. In the zero temperature limit, we have a Poissonian Fano factor, $F=1$, when the tunneling of subsequent electrons is uncorrelated. The Fano factor is sub-Poissonian, $F<1$, if the electronic-transport occurs in an anti-correlated way, while we obtain a super-Poissonian Fano factor, $F>1$, for correlated electron tunneling.
\section{Enhancement of the Fano factor from repulsive interaction \label{sec:Enhancement of the Fano factor from repulsive interaction}}

In this section, we are going to show that a many-state model including neutral plasmon excitations is able  to capture the features of a strongly interacting interferometer. As a consequence of the non-equilibrium excitation of  neutralon modes, the Fano factor is enhanced for a sufficiently strong edge-edge coupling.\\
In the following, we consider a symmetric interferometer with $\gamma_L = \gamma_R = \gamma$, and the bias voltage is applied symmetrically $V_{L/R} = \pm V/2$. We assume that the value of $\gamma$ is independent of the filling factor, and we determine its value by comparing the conductance of a non-interacting FPI ($\Lambda=0$, or equivalently $\nu=1$) with the conductance of a FPI in the coherent tunneling regime (see \cref{sec:Coherent tunneling}). However, we would like to point out that ultimately the magnitude 
$\gamma$ of the tunneling rate is not important because the Fano factor is independent of it, since both the current and the noise are linear in $\gamma$.

\begin{figure}
	\centering
	\includegraphics[scale=0.37]{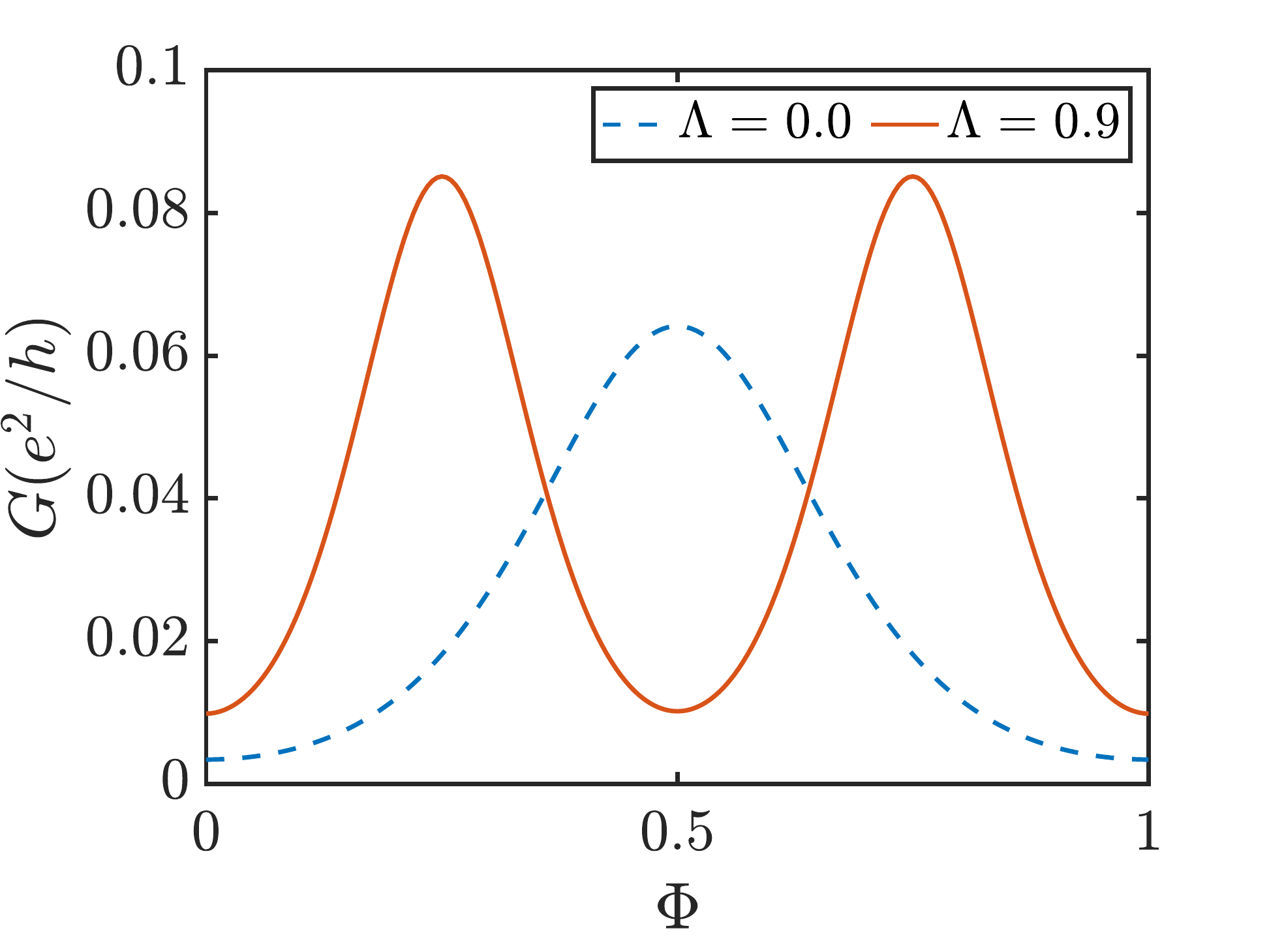}
	\caption{Conductance as a function of the flux for edge-edge interaction strength $\Lambda = 0$ (dashed line) and $\Lambda = 0.9$ (solid line). For strong interactions, the magnetic field periodicity is halved, similar to the behavior found for an FPI in the open limit   \cite{frigeri2019sub}. The temperature is $\beta E_C=10$ and $\gamma= 0.0513$.}\label{fig:conductance filling2}
\end{figure}

Using the rate equation formalism described in the previous section, we first compute the average current and the conductance. In \cref{fig:conductance filling2}, we plot the conductance as a function of the magnetic flux for $\Lambda=0.0$ and $\Lambda=0.9$. In the absence of interactions, $\Lambda=0$, the system is in the Aharonov-Bohm (AB) regime \cite{frigeri2019sub,choi2015robust,sivan2016observation,ABvsCD,halperin2011theory}, and the flux periodicity of the conductance is one flux quantum, $\Delta\Phi=1$. The strong edge-edge interaction places the interferometer in the so-called AB$^\prime$ regime \cite{frigeri2019sub,choi2015robust,manfra2019}, where we observe a conductance with a flux periodicity equal to half the magnetic flux quantum, $\Delta \Phi=1/2$. The fact that we obtain the halving of the oscillation period for an interferometer in the both the closed and in the open limit  \cite{frigeri2019sub} is due to the fact that the closed limit conductance can be expressed as  a Fourier series in the magnetic flux, and that the open limit contributes the leading periodicity in this Fourier expansion. 
We want to remark that the plasmon excitations are not fundamental to describe magnetic flux dependence of the conductance,  and that they are not responsible for the period halving, since they do not couple to the magnetic flux as it can be seen from \cref{eq:energy filling2}.

\begin{figure}[h]
	\centering
	\includegraphics[scale=0.37]{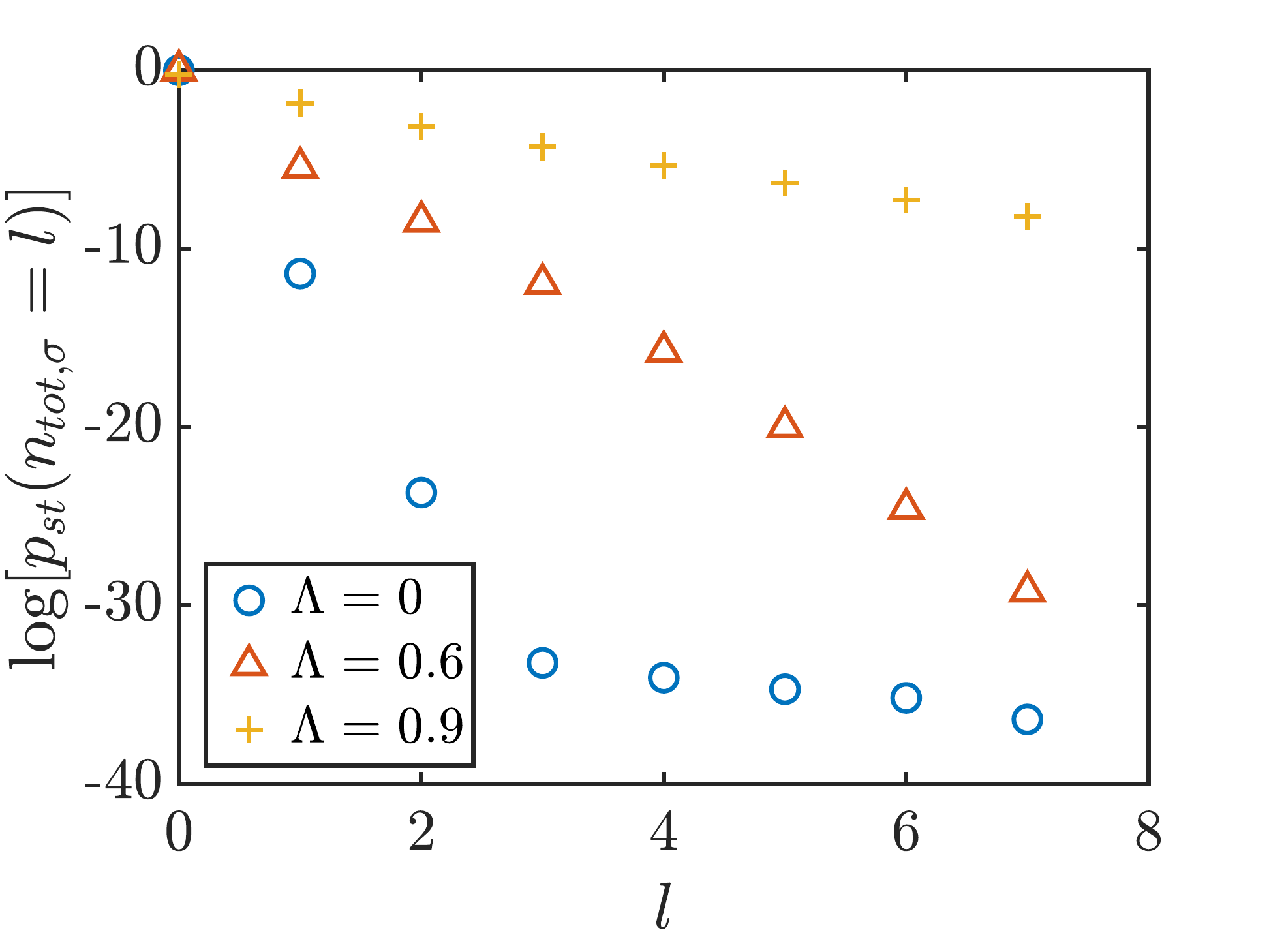}
	\caption{Probability (in log-scale) to have $l$ neutral plasmon excitations in the system, defined in \cref{eq:prob n_sigma tot}, for different values of the edge-edge interaction strength $\Lambda$. A strong intra-edge coupling favors the creation of the neutral plasmon excitations.The magnetic flux is set to $\Phi=0$, the temperature is $\beta E_C=20$ and the bias-voltage $eV/E_C=0.5$.}\label{fig:prob neutral plasmons}
\end{figure}

\begin{figure}[h]
	\centering
	\subfloat{\label{fig:noise filling2}\includegraphics[scale=0.37]{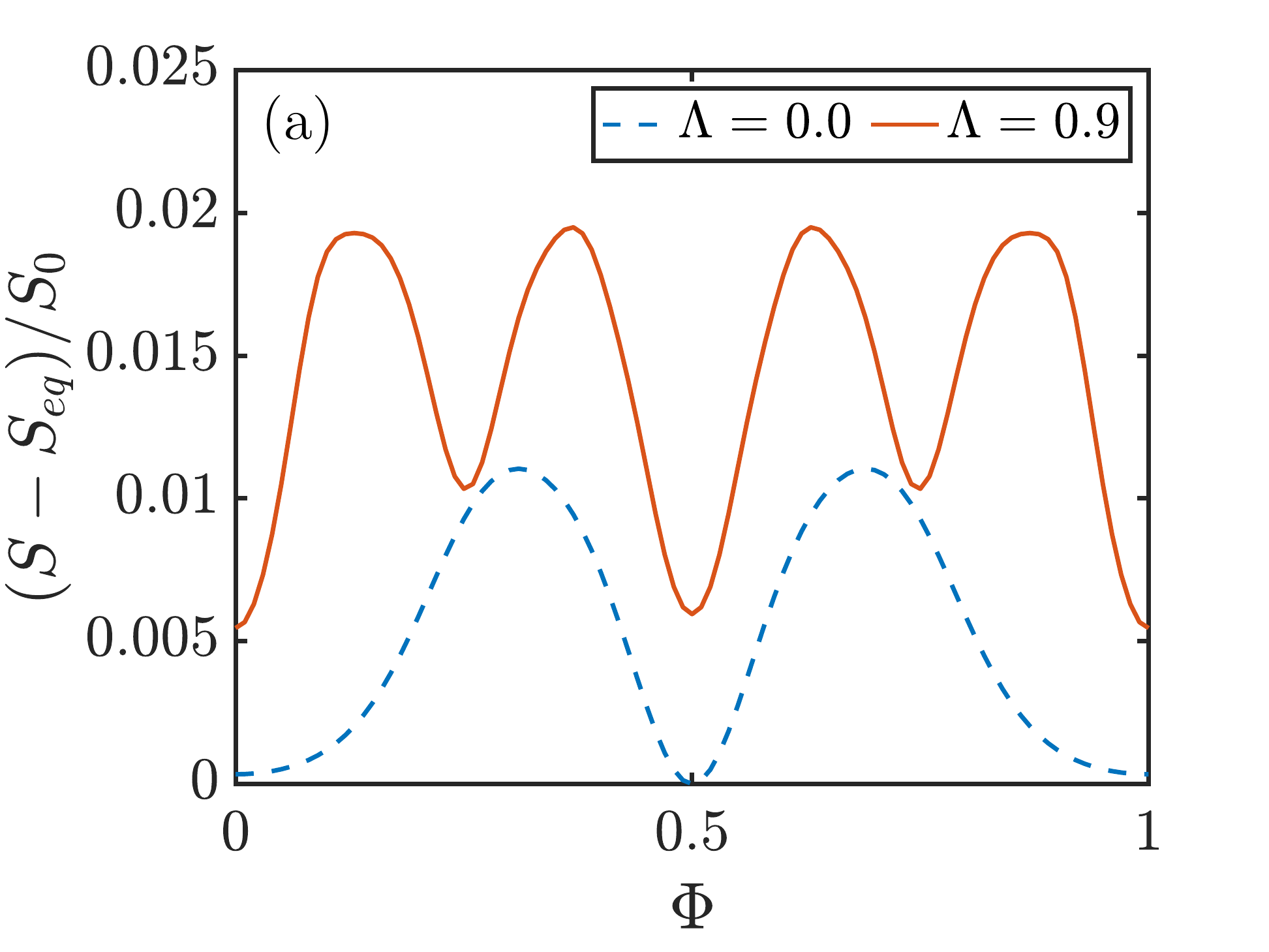}}\\
	\subfloat{\label{fig:Fano filling2}\includegraphics[scale=0.37]{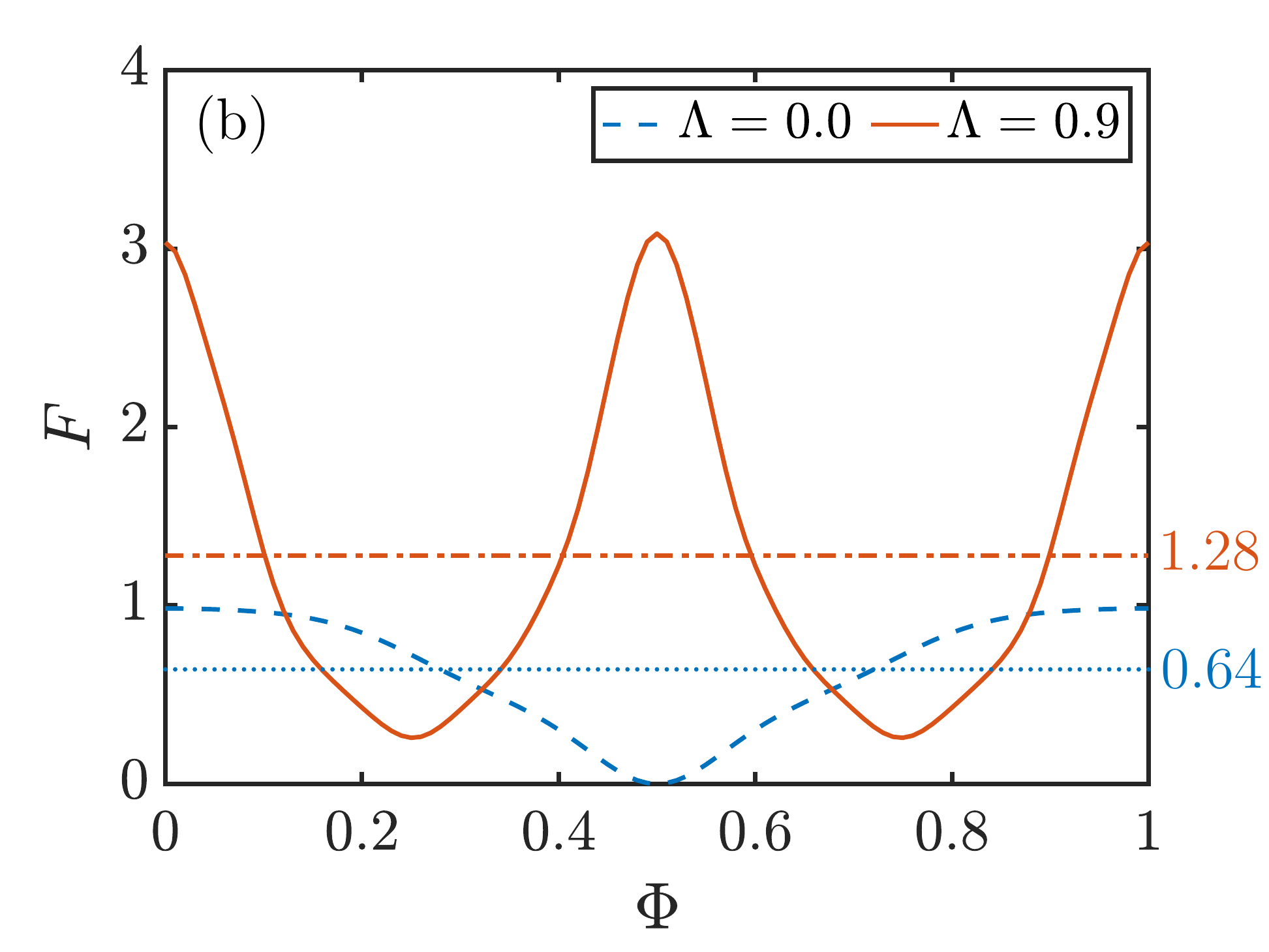}}
	\caption{(a) Excess noise $S-S_{eq}$, normalized by the reference noise $S_0 = 2e^2 E_C/h$, as a function of the flux for edge-edge interaction strength $\Lambda = 0$ (dashed line) and $\Lambda = 0.9$ (solid line). The parameters are $\beta E_C=20$, $eV/E_C=0.5$ and $\gamma= 0.0513$. (b) Fano factor $F$ as a function of the flux in the non-interacting limit ($\Lambda=0$, dashed line) and in the strongly interacting limit ($\Lambda=0.9$, solid line). The Fano factor is significantly enhanced in the presence of strong inter-edge coupling. The dotted (dashed-dotted) line represents the average value of the Fano factor for $\Lambda = 0$ ($\Lambda = 0.9$), as defined in \cref{eq:average Fano}. We set the temperature $\beta E_C=20$ and bias-voltage $eV/E_C=0.5$.}\label{fig:Noise-Fano filling2}
\end{figure}

However, we will  see momentarily that the neutralon excitations are a key ingredient for the enhancement of current fluctuations and the Fano factor. First, we look at the probability of neutral plasmon excitations in the system. Given the neutral plasmons configuration $\{n\}_\sigma = \{n_{1,\sigma}, n_{2,\sigma}, \dots\}$, we define the total number of neutral plasmon $n_{tot,\sigma}$ as
\begin{equation}
n_{tot,\sigma}  = \sum_{m=1}^{\infty} n_{m,\sigma}.
\end{equation}
The probability to have $l$ neutral plasmons in the dot is defined as
\begin{equation}\label{eq:prob n_sigma tot}
p_{st}(n_{tot,\sigma} = l) = \sum_{\substack{N_1, N_2 \\ \{n\}_\rho}} \sideset{}{'}\sum_{\{n\}_\sigma} p_{st}(N_1, N_2, \{n\}_\rho, \{n\}_\sigma),
\end{equation}
where $p_{st}(N_1, N_2, \{n\}_\rho, \{n\}_\sigma)$ is obtained from the master equation, and the primed sum is constrained to those neutral plasmons configurations $\{n\}_\sigma$ satisfying $n_{tot,\sigma} = l$.
In \cref{fig:prob neutral plasmons}, we report the probability defined in \cref{eq:prob n_sigma tot} for different values of the edge-edge interaction $\Lambda$ at $\Phi=0$, $\beta E_C=20$ and $eV/E_C=0.5$. It is evident that a strong edge-edge coupling make it easier to create the neutral plasmons. Indeed, a smaller amount of energy is required to excite them because the energy gap for the neutral plasmons $E_{\sigma}$ decreases with the strength of the edge-edge interaction $\Lambda$, as can be seen from \cref{eq:epsilon rho/sigma}. While for an intermediate interaction strength $\Lambda = 0.6$ the excitation probability decreases quickly with increasing excitation number $l$,  for a large interaction strength $\Lambda = 0.9$ a large number of neutral plasmons can be excited, and a 
many-state model is necessary to correctly describe the  properties of a strongly interacting interferometer. 
\\
In \cref{fig:Noise-Fano filling2}, we report the excess noise and the Fano factor as a function of the flux for a non-interacting, $\Lambda=0$, and a strongly interacting FPI, $\Lambda=0.9$. The excess noise is an oscillatory function of the flux, and we observe that the current noise is more pronounced for  stronger inter-edge interactions. Indeed, both the maximum and minimum values of the excess noise for $\Lambda=0.9$ are larger than the respective values in the non-interacting case. 
At $\Lambda=0$, the excess noise vanishes for $\Phi=1/2$, since for this value of flux the  total noise is entirely due to thermal noise, $S=S_{eq}$.
On the contrary, the excess noise never vanishes, $S >  S_{eq}$, in the strongly interacting case, and this is a signature of the participation of the neutral plasmons to the transport. 
Both the excess noise and the Fano factor have the same flux periodicity as the conductance. By comparing  \cref{fig:conductance filling2} with \cref{fig:Fano filling2}, we see that when the conductance increases, in general  the Fano factor decreases and a maximum/minimum in $F$ corresponds to a minimum/maximum in $G$. 
An increase of the conductance implies that  electrons are more easily transported through the system. 

\begin{figure}
	\centering
	\includegraphics[scale=0.37]{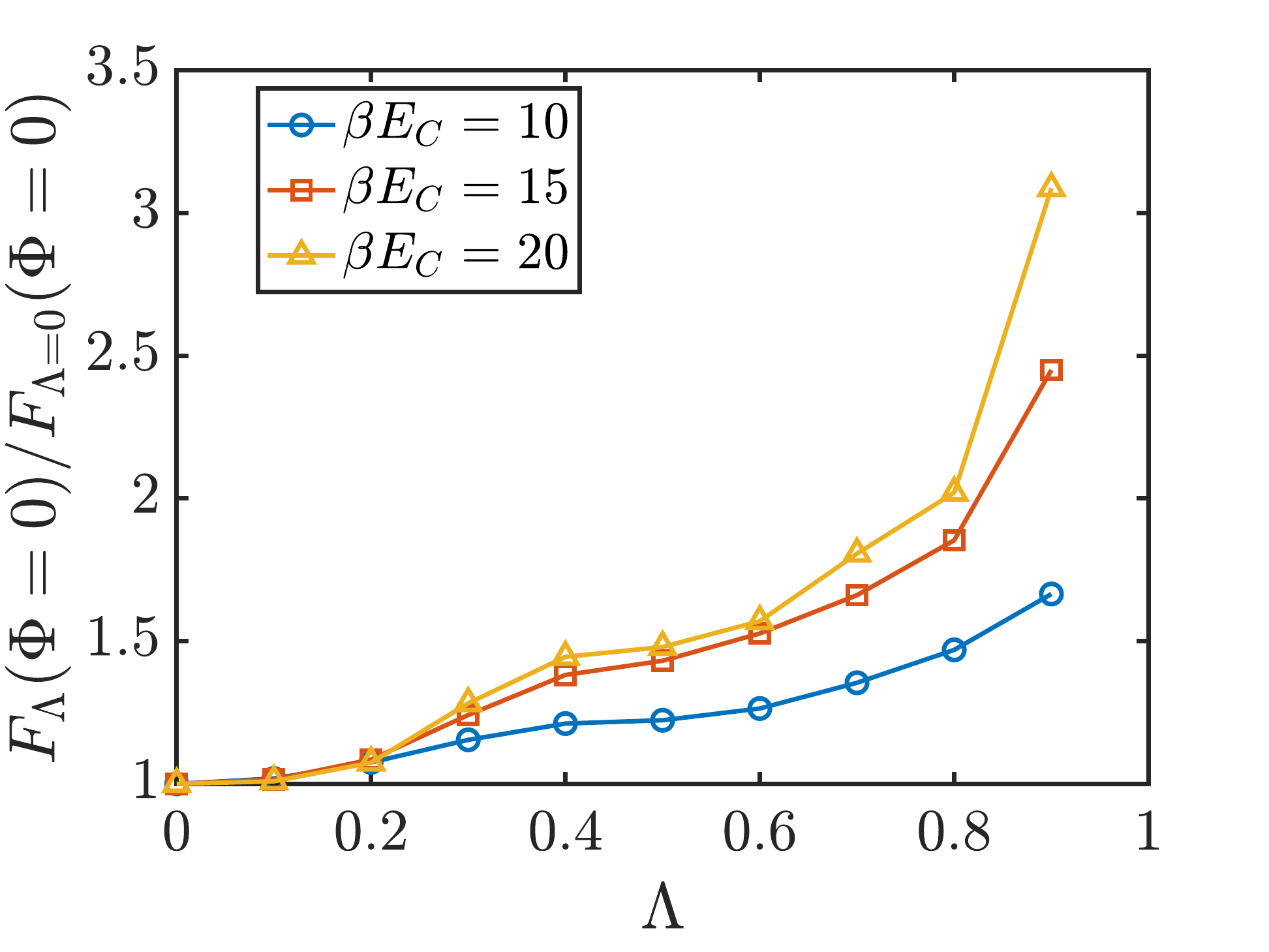}
	\caption{Ratio between the Fano factor $F_\Lambda$ in the presence of interaction and the non-interacting case $F_{\Lambda=0}$ at flux $\Phi = 0$ for different values of $\Lambda$ and three different temperatures ($\beta E_C = 10, 15, 20$). The enhancement of the Fano factor is more pronounced for stronger edge-edge coupling $\Lambda$ and lower temperature.}\label{fig:enhancement Max Fano}
\end{figure}

When one electron enters into the dot, it is necessary to wait for that electron to leave the dot before adding another electron, as a consequence of the Pauli exclusion principle. Therefore, the transfer of two subsequent electrons is anti-correlated,  and  the Fano factor is suppressed with respect to the Poissonian case of non-interacting electrons.
However, the transfer of subsequent electrons is almost uncorrelated at $\Phi=0$ in the absence of inter-edge interactions, $\Lambda=0$, because in this case transport is limited by the exponentially suppressed rate for electrons entering the dot in the limit 
$\beta E_C \gg 1$, while electrons leave the dot almost instantaneously. As a consequence, for  $\Phi=0$ and $\Lambda=0$ we 
obtain a Fano factor  $F\approx 1$. By increasing $\Phi$, the effect of Pauli blockade becomes more prominent, and this gives a sub-Poissonian Fano factor, $F<1$. At  the energetically degenerate point at $\Phi=0.5$, the noise is totally thermal and hence the Fano factor vanishes, $F=0$.
On the other hand, the participation of neutralons in electron transport in the strongly interacting limit implies an enhancement of the Fano factor, as  can be seen from \cref{fig:Fano filling2}. Indeed, the maximum of $F$ at $\Lambda=0.9$ is three times higher than the maximum of $F$ at $\Lambda=0$, indicating a significant correlation between subsequent tunneling events. In addition, due to the excitation of neutral plasmons implies,  the minimum of $F$ occurs at a finite value. 
Hence, we have two competing mechanisms affecting the Fano factor: the Pauli exclusion principle in combination with the charging energy of the dot tends to reduce it, while an upward correction is caused by the involvement of the neutralons in  electron transport. We also indicate in \cref{fig:Fano filling2} the average value of $F$ with respect to the flux, defined as
%
\begin{equation}\label{eq:average Fano}
\langle F \rangle_\Phi = \int_{0}^{1} d\Phi~F \left(\Phi\right).
\end{equation}
To quantify the enhancement of the Fano factor, we compare in \cref{fig:enhancement Max Fano} the Fano factor $F_\Lambda$ in presence of interaction with the non-interacting case $F_{\Lambda=0}$ at $\Phi = 0$ for different values of $\Lambda$. As in \cref{fig:enhancement_intro}, the Fano factor at $\Phi=0$ is more strongly enhanced for larger inter-edge coupling $\Lambda$, and at lower temperature. 
To summarize, we have seen that a strong edge-edge interaction favors the excitation of neutral plasmons, and also leads to an   enhancement of the Fano factor. In the next section, we will establish a link between the presence of neutralon excitations and an enhancement of the Fano factor. 
%
\section{Electron attraction mediated by neutral plasmons \label{sec:Electron attraction mediated by plasmon excitations}}
In this section  we want to show that it is possible for  electrons passing through the interfering edge to attract each other via the exchange of neutral plasmon excitations in the presence of a sufficiently strong repulsive  interaction. The formation of electron bunches explains  the  enhancement of the Fano factor discussed in the previous section.

 We now argue that  for intermediate values of the interaction it is sufficient to consider  a three-state model with charge states $N_1 = 0,1$ and the lowest energy neutral excitation $\{n\}_\sigma = \{1,0,\dots\}_\sigma \equiv \{1\}_\sigma$, defined by the basis states
\begin{equation}
\lvert 0, \{0\}_\sigma \rangle, \qquad \lvert 1, \{0\}_\sigma \rangle, \qquad \lvert 0, \{1\}_\sigma \rangle \ .
\end{equation}
Here,  we have a fixed $N_2 = 0$, and do not consider excitation of charged plasmon. In \cref{fig:Fano 3 states},  the Fano factor as obtained from the rate equation formalism is plotted as a function of  flux for the  interaction strength $\Lambda=0.6$, temperature $\beta E_C = 20$ and voltage $eV/E_C = 0.5$. By comparing the result of a two-state model (without any plasmon excitations), the three-state model defined above, and the full  model, we see that it is necessary to take into account  plasmon excitations in order to correctly model the properties of an FPI with the chosen parameters. In addition,  the three-state model captures a large part of the  Fano factor enhancement. 
Therefore, the most relevant transport channels are the ones described in  \cref{eq:seq,eq:neutr}. 
As explained in \cref{sec:Introduction}, the exchange of a neutral plasmon makes the tunneling of subsequent electrons correlated and this produces the enhancement of the Fano factor.
We note that the exact Fano factor in \cref{fig:Fano 3 states} does not yet have the AB$^\prime$ periodicity $\Delta\Phi = 1/2$. Indeed, we see that at $\Lambda = 0.6$ a sub-leading AB component with periodicity $\Delta\Phi = 1$ is present in the Fourier spectrum of $F$, in addition to  the dominant AB$^\prime$ component, and it cannot be neglected  \cite{frigeri2019sub}. By increasing the value of $\Lambda$, the AB component becomes less important. However, the three-state model is inapplicable if we increase the value of $\Lambda$, because it becomes easier to excite multiple plasmon states and so a many-states model would be necessary to describe the properties of the interferometer.
\begin{figure}
	\centering
	\includegraphics[scale=0.37]{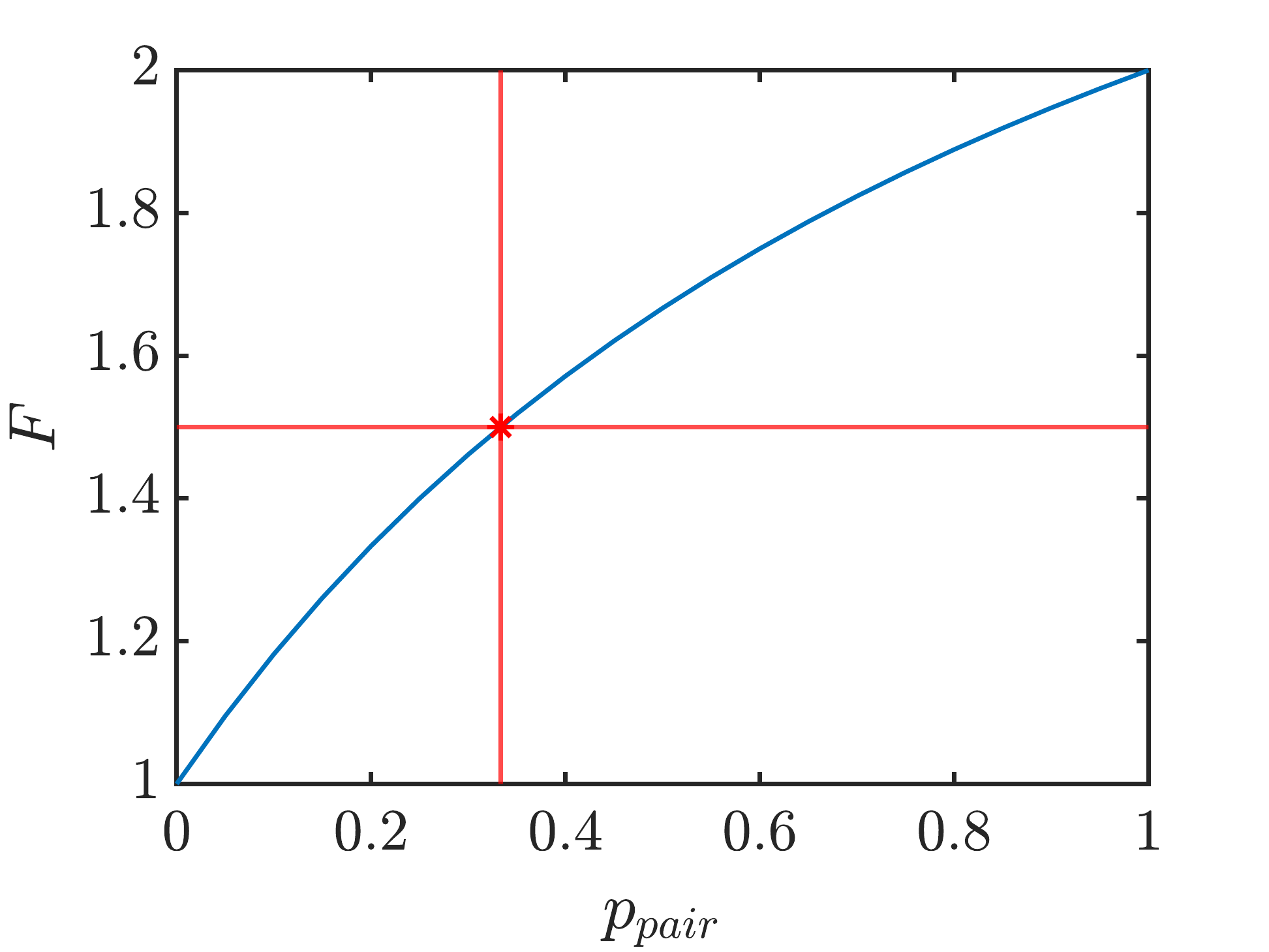}
	\caption{Fano factor as a function of the probability for an electron pair to tunnel $p_{\rm pair}$ for the model described by \cref{eq:p N_el,eq:p N_events,eq:conditional N_el given N_events}, in which individual electrons and pairs tunneling coexist. The point indicated with a red star at $p_{\rm pair} =  p_{\rm single}/2 = 1/3$ describes the 3-state model in \cref{fig:Fano 3 states} at $\Phi = 0$, as it can be seen by looking at the transition rates in the tree diagram of \cref{fig:transition rates}. The Fano factor at $p_{\rm pair}=1/3$ is $F=1.5$, in very good agreement with the the enhancement found in \cref{fig:enhancement Max Fano} at $\Lambda=0.6$.}\label{fig:Fano pair}
\end{figure}

Finally, we want to confirm that the enhancement of the Fano factor can be interpreted in terms of single electrons and electron pairs tunneling. 
The probability that $N_{\rm el}$ electrons tunnel through the system is given by
\begin{equation}\label{eq:p N_el}
p\left(N_{\rm el} \right) = \sum_{N_{\rm events} = N_{\rm el}/2}^{N_{\rm el}} p\left(N_{\rm el} \,\lvert\, N_{\rm events} \right) p\left(N_{\rm events} \right),
\end{equation}
where $N_{\rm events}$ is the number of tunneling events (either tunneling of single electrons or of electron pairs), $p\left(N_{\rm events} \right)$ is the corresponding probability and $p\left(N_{\rm el} \,\lvert\, N_{\rm events} \right)$ is the conditional probability for a total of $N_{\rm el}$ electrons to tunnel,  given $N_{\rm events}$ tunneling events. The lower bound of the summation in \cref{eq:p N_el} corresponds to the case in which only electron pairs tunnel, while the upper bound corresponds to a situation in which   only individual electrons tunnel. We assume that $p\left(N_{\rm events} \right)$ is given by a Poisson distribution expressed in terms of the average number of tunneling events $\langle N_{\rm events}\rangle$
\begin{equation}\label{eq:p N_events}
p\left(N_{\rm events} \right) = \frac{\langle N_{\rm events}\rangle^{N_{\rm events}}}{N_{\rm events}!} e^{-\langle N_{\rm events}\rangle}.
\end{equation}
Given $N_{\rm events}$ tunneling events, we have $k$ single electrons and $(N_{\rm events}-k)$ pairs tunneling. Accordingly, the number of tunneled electrons is given by
\begin{equation}
N_{\rm el} = k+2(N_{\rm events}-k) = 2 N_{\rm events} -k,
\end{equation}
and the conditional probability is a binomial distribution
\begin{equation}\label{eq:conditional N_el given N_events}
p\left(N_{\rm el} \lvert N_{\rm events} \right) = \binom{N_{\rm events}}{2N_{\rm events}-N_{\rm el}} p_{\rm single}^{2N_{\rm events}-N_{\rm el}} 
p_{\rm pair}^{N_{\rm el}-N_{\rm events}}.
\end{equation}
By combining \cref{eq:p N_el,eq:p N_events,eq:conditional N_el given N_events}, we can obtain the average number of tunneled electrons $\langle N_{\rm el} \rangle$, the variance $\Var\left(N_{\rm el}\right)$
and the Fano factor $F = \Var\left(N_{\rm el}\right)/ \langle N_{\rm el} \rangle$. In \cref{fig:Fano pair}, we plot the Fano factor obtained from \cref{eq:p N_el,eq:p N_events,eq:conditional N_el given N_events} as a function of $p_{\rm pair}$. 
In particular, we have $F=1.5$ (red star in \cref{fig:Fano pair}) at $p_{\rm pair} = p_{\rm single}/2 = 1/3$, representing the interferometer for the parameters $\Lambda=0.6$, $\beta E_C = 20$, $eV/E_C = 0.5$, $\Phi = 0$ (compare the transition rates in \cref{fig:transition rates}). This value is in very good agreement with actual enhancement $F_{\Lambda}(\Phi=0)/F_{\Lambda=0}(\Phi=0) = 1.57$ found in \cref{fig:enhancement Max Fano}, implying that   our results can indeed be interpreted in terms of single and electron pairs tunneling.
\section{Comparison between Master equation and scattering formalism \label{sec:Coherent tunneling}}

The goal of this section is twofold. First, we analyze under what conditions results obtained from the master equation formalism are applicable to the description of  Fabry-P\'{e}rot interferometers with a finite QPC transparency. Second, we explain that in the absence of dephasing it is not possible to extract an effective charge from excess noise computed in the master equation formalism. For this reason, we have presented 
our results in the preceeding section in terms of a relative enhancement of the Fano factor rather than in terms of an effective interfering charge. 

A key ingredient in interferometry is phase coherence. Therefore, it is natural to ask  under which condition the results from the master equation approach are equivalent to those obtained from a scattering formalism describing  coherent tunneling of electrons.
To this end, we here compare  observables characterizing electronic-transport of an FPI at filling factor $\nu=1$
obtained with the master equation (ME) with results obtained via 
 the scattering formalism (SF) \cite{Buttiker-SF1,ButtikerSF2,blanter2000shot}. 

\begin{figure}
	\centering
	\subfloat{\label{fig:conductance filling 1}\includegraphics[scale=0.37]{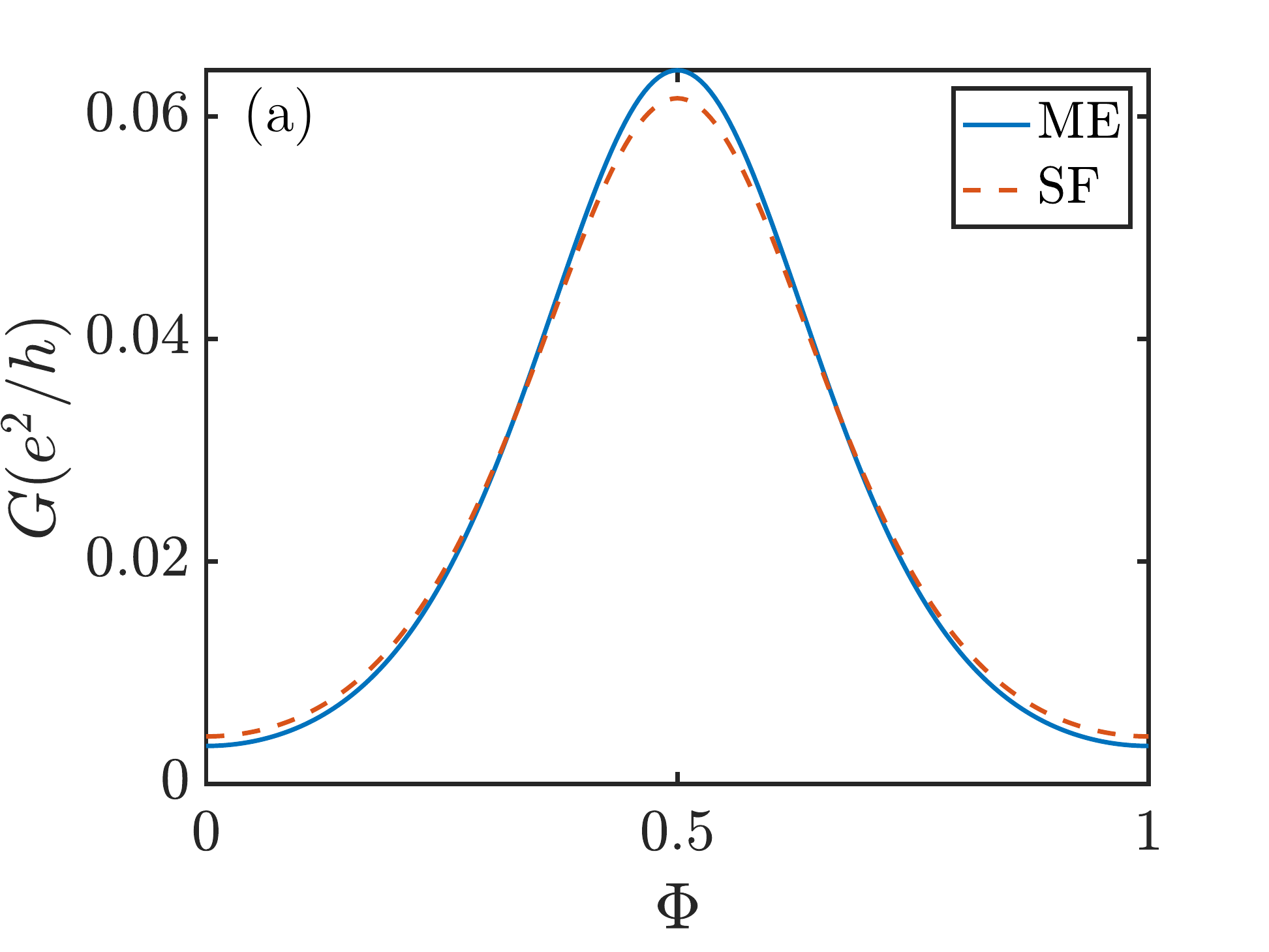}}\\
	\subfloat{\label{fig:error filling 1}\includegraphics[scale=0.37]{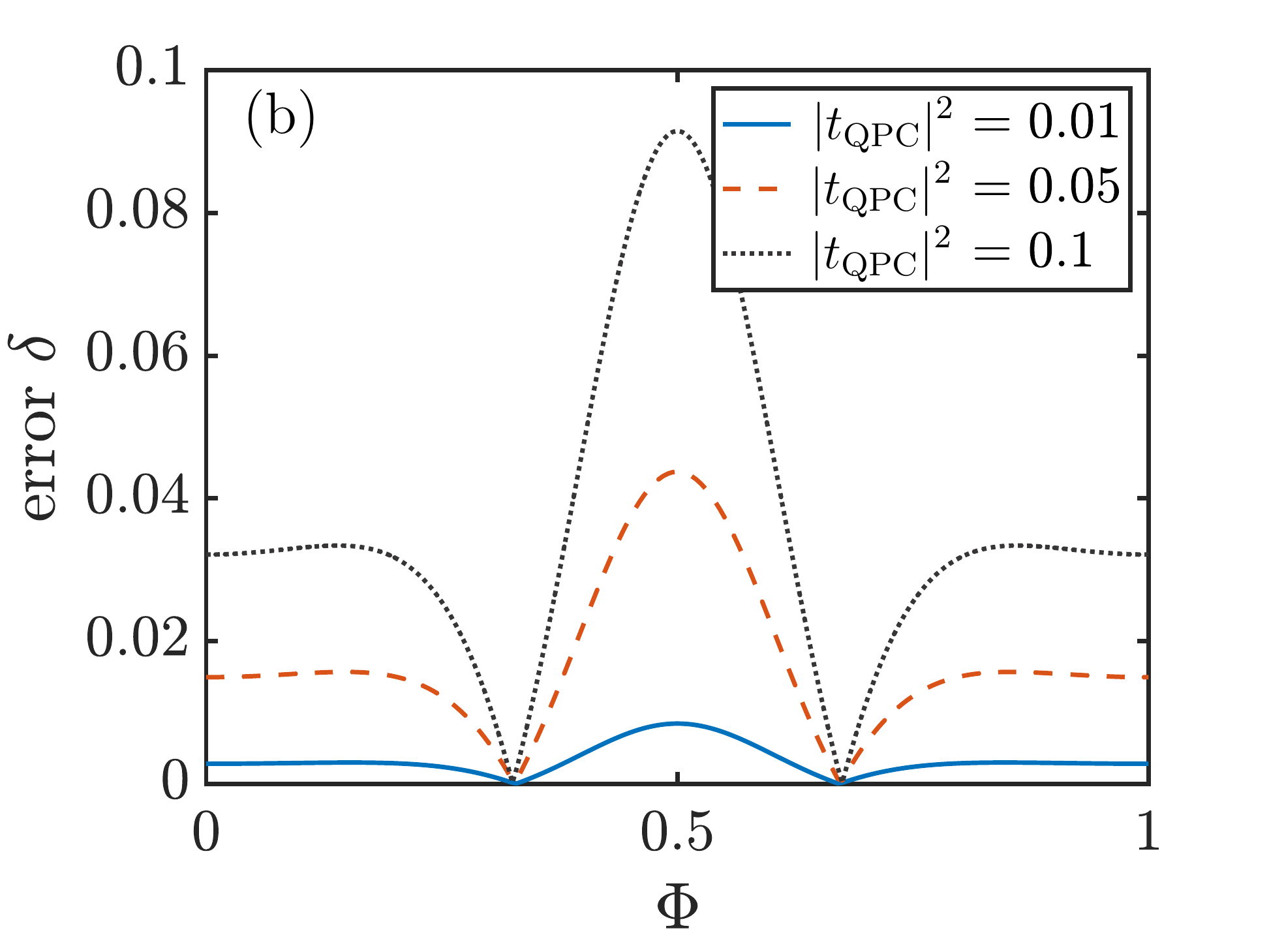}}
	\caption{(a) Conductance at filling factor $\nu=1$ (equivalent to $\Lambda=0$) as a function of the flux calculated from the master equation (ME, solid line) and the scattering formalism (SF, dashed line). The temperature is chosen as $\beta E_C = 10$, and the QPC transparency is $\lvert t_{\rm QPC} \rvert^2 = 0.05$ in the SF. By requiring that the value of the flux-averaged conductance is the same for the SF and the ME, we can determine the bare tunneling rate entering in the ME to be $h\gamma/E_C = 0.0513$. (b) Error $\delta$, defined in \cref{eq:error delta}, as a function of the flux at $\beta E_C = 10$ for different values of the QPC's transparency $\lvert t_{\rm QPC} \rvert^2$, such that the condition in \cref{eq:condition} is fulfilled. The error decreases when the interferometer is more closed.}
	\label{fig:MEvsSF}
\end{figure}

The fundamental quantity that we need to compute in the SF is the transmission probability $\mathcal{T}$. In a round trip along the interference cell, the electron picks up the phase
\begin{equation}\label{eq:theta FPI}
	\theta(E) = 2\pi \left(\frac{E}{E_C} + \Phi\right),
\end{equation}
where the first part is a dynamical phase and the second contribution is the Aharonov-Bohm phase \cite{AB_effect}. When summing over all possible numbers of windings around the interference cell, the transmission probability of the FPI is found to be 
\begin{equation}\label{eq:transmission probability FPI}
	\mathcal{T} (E) = \frac{\lvert t_L \rvert^2 \lvert t_R \rvert^2}{\lvert 1 + \lvert r_L \rvert \lvert r_R \rvert e^{2\pi i (E/E_C + \Phi)}\rvert^2} \ ,
\end{equation}
with transmission (reflection) amplitude $t_l$ ($r_l$) at the $l$-th QPC. 
When deriving \cref{eq:transmission probability FPI}, we assumed that $\sgn(r_L  r_R) = -1$ such that the conductance obtained with the SF has a maximum at $\Phi = 0.5$, like the conductance computed with the ME at $\Lambda = 0$ (see \cref{fig:MEvsSF}).
In the weak tunneling limit $\lvert t_L \rvert$, $\lvert t_R \rvert \ll 1$, the transmission probability is sharply peaked around the discrete energy levels $E_n$, such that $\theta(E_n) = \pi(1+2n)$, and it can be  approximated by the Breit-Wigner formula \cite{ihn2010semiconductor}
\begin{equation} \label{eq:breitwigner}
	\mathcal{T} (E) \simeq \sum_n \frac{\Gamma_{{\rm FPI}, L} \Gamma_{{\rm FPI},R}}{(\Gamma_{{\rm FPI}}/2)^2 + (E + E_C\Phi-E_n)^2} \ ,
\end{equation}
with the total width of the resonant level given by 
\begin{equation}\label{eq:width FPI}
	\Gamma_{{\rm FPI}} = \Gamma_{{\rm FPI},L} + \Gamma_{{\rm FPI},R}\ ,
\end{equation}
obtained from the partial widths 
\begin{equation}\label{eq:partial width}
	\Gamma_{{\rm FPI},L/R} = \frac{E_C}{2\pi} \lvert t_{L/R} \rvert^2\ .
\end{equation}

Approximating the transmission probability \cref{eq:transmission probability FPI} by \cref{eq:breitwigner} becomes 
exact in the limit $\Gamma_{{\rm FPI}}/E_C \to 0$. 
We can obtain from \cref{eq:transmission probability FPI} the current \cite{blanter2000shot}
\begin{equation}\label{eq:current SF}
	I = \frac{e}{2\pi} \int dE~\mathcal{T}(E) \left[f_L(E)-f_R(E)\right] \ ,
\end{equation}
and subsequently the two terminal conductance
\begin{equation}\label{eq:G SF}
	G = \frac{e^2}{2\pi} \int dE~\mathcal{T}(E) \left(-\frac{\partial f}{\partial E}\right).
\end{equation}
Finally, the noise is given by  \cite{blanter2000shot}
\begin{align}
	S = \frac{e^2}{\pi} \int dE \big\{\mathcal{T}(E) \left[f_L (1-f_L) + f_R (1-f_R)\right] \nonumber\\
	+ \mathcal{T}(E) \left[1-\mathcal{T}(E)\right] (f_L-f_R)^2\big\} \ .\label{eq:noise SF}
\end{align}
The Fano factor is then obtained from \cref{eq:current SF,eq:noise SF} and by using the definition \cref{eq:Fano factor ME}.

\begin{figure}
	\centering
	\subfloat{\label{fig:noise filling 1}\includegraphics[scale=0.37]{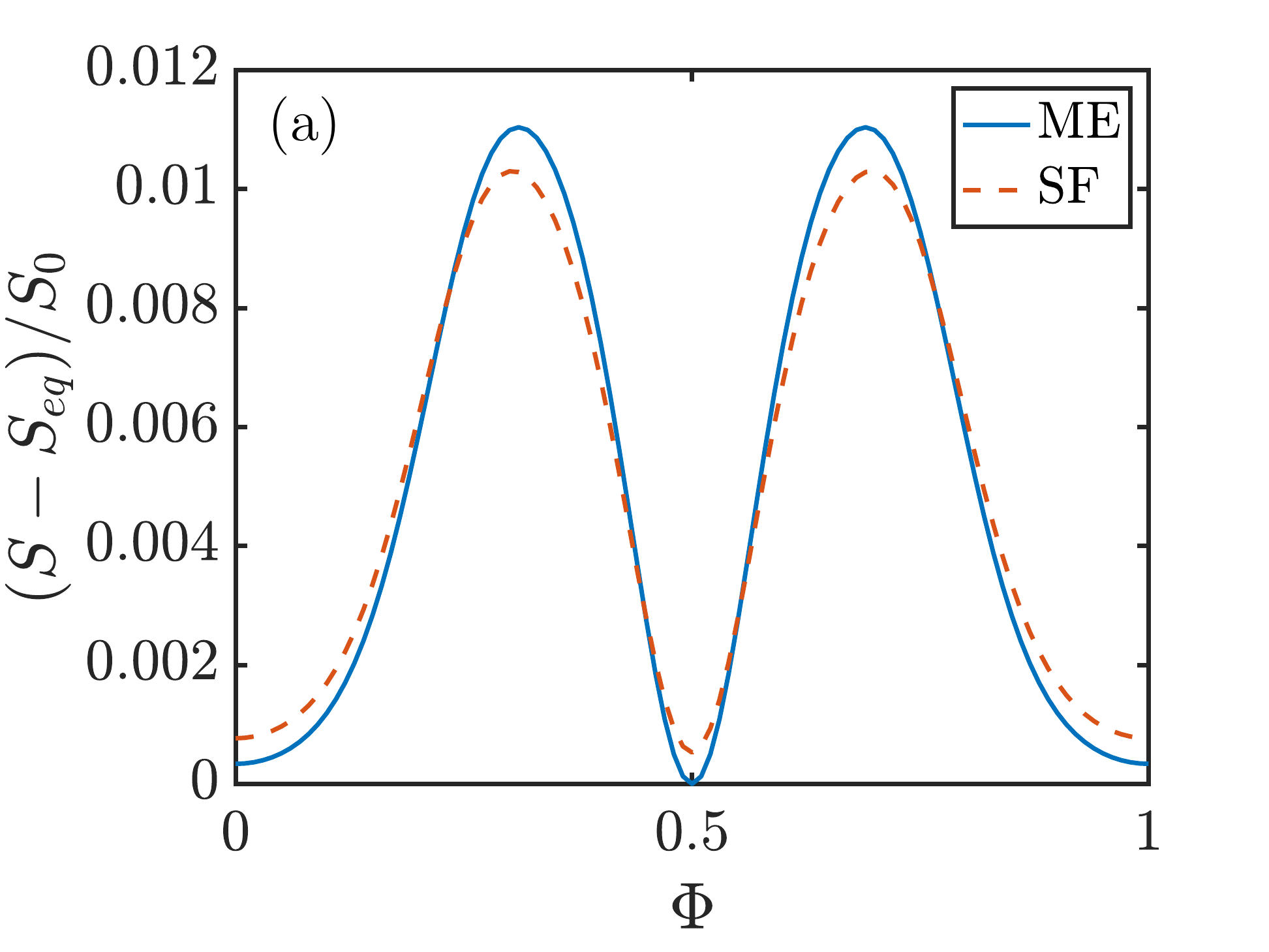}}\\
	\subfloat{\label{fig:Fano filling 1}\includegraphics[scale=0.37]{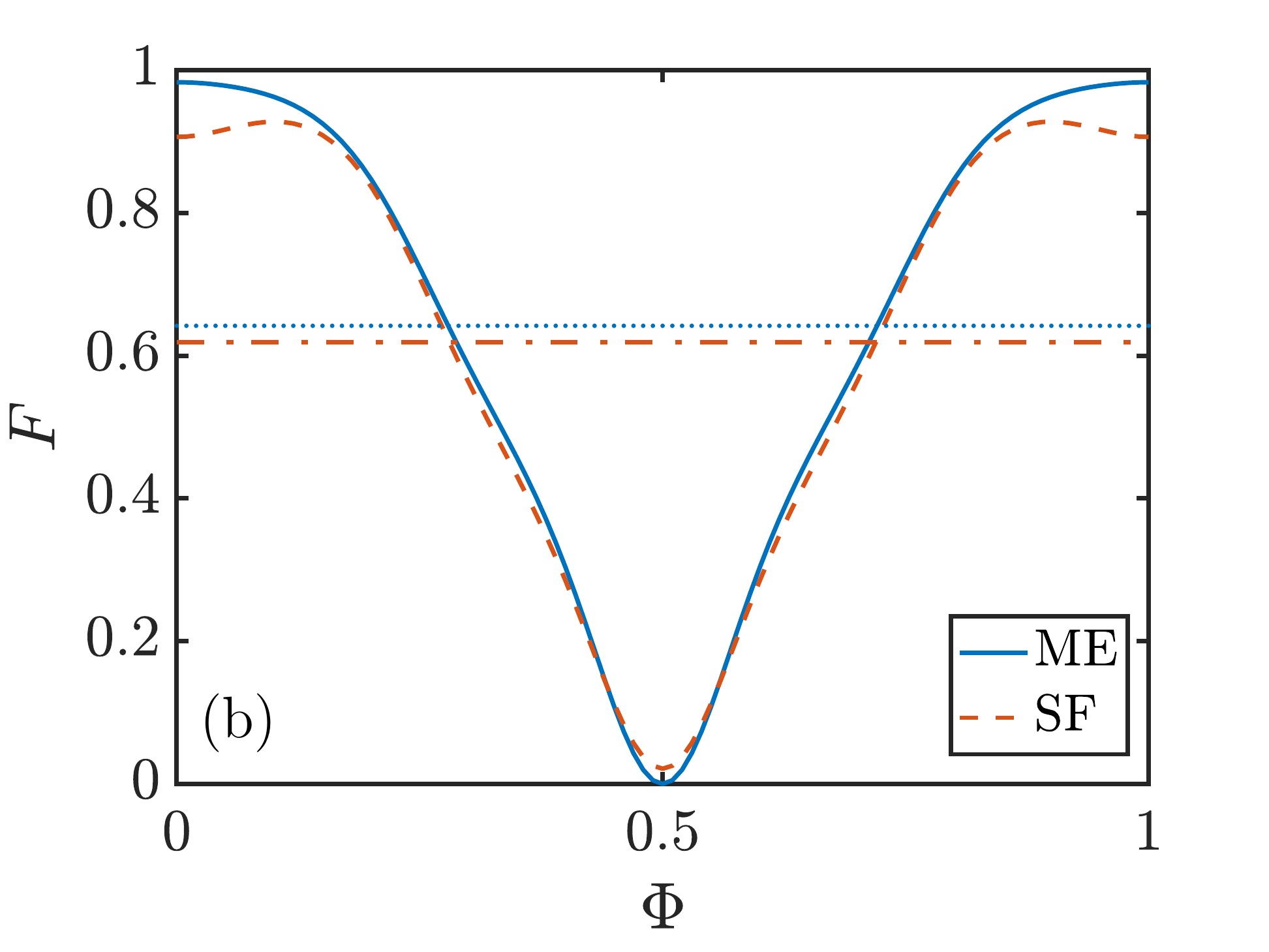}}
	\caption{(a) Excess noise $S-S_{eq}$ divided by the reference noise $S_0 \equiv 2 e^2 E_C/h$ at filling factor $\nu=1$ as a function of the flux $\Phi$ obtained from the master equation (ME) and the scattering formalism (SF). Parameters are chosen as $\beta E_C = 20$, bias-voltage $eV/E_C=0.5$, $\lvert t_{\rm QPC} \rvert^2 = 0.05$ and $h\gamma/E_C = 0.0513$.  (b) Fano factor at filling factor $\nu=1$ as a function of the flux $\Phi$ obtained from the master equation (ME) and the scattering formalism (SF). The dotted/dashed-dotted line represents the average value of the Fano factor, defined in \cref{eq:average Fano}, for the ME/SF. Parameters are chosen as $\beta E_C = 20$, bias-voltage $eV/E_C=0.5$ and $\lvert t_{\rm QPC} \rvert^2 = 0.05$.}
	\label{fig:Noise-Fano filling 1}
\end{figure}

In the following, we assume the FPI to be symmetric, by setting $\lvert t_L \rvert = \lvert t_R \rvert \equiv \lvert t_{\rm QPC} \rvert$ for the SF in \cref{eq:transmission probability FPI}.  The symmetry of the FPI is reflected in the ME by letting $\gamma_L = \gamma_R \equiv \gamma$ in \cref{eq:addition rate filling2,eq:removing rate filling2}. Moreover, we assume that the voltage bias is applied symmetrically, $V_{L/R} = \pm V/2$. 
One important difference between the SF and the ME is that the intrinsic width of the resonant level is neglected in the ME. 
In order to compare results between the two formalisms, we hence  consider the limit in which the temperature $k_BT$ is much bigger than the intrinsic width of the resonant level $\Gamma_{\rm FPI}$. By using \cref{eq:partial width,eq:width FPI} and the symmetry of the FPI, this condition is expressed as
\begin{equation}\label{eq:condition}
\beta E_C \ll \frac{\pi}{\lvert t_{\rm QPC} \rvert^2}.
\end{equation}
In \cref{fig:conductance filling 1}, we show the conductance as a function of the flux calculated from the master equation  $\nu=1$ (ME, solid line) and the scattering formalism in \cref{eq:G SF} (SF, dashed line), under the condition in \cref{eq:condition}. The bare tunneling rate $\gamma$ entering in the ME is determined by requiring that the average value of the conductance is the same for the ME and the SF. The conductance is periodic in the flux with a  period of one flux quantum $\Delta\Phi = 1$, as expected from the Aharonov-Bohm effect \cite{choi2015robust,halperin2011theory,frigeri2019sub}. We see that at integer values of the flux $\Phi = n \in \mathbb{Z}$ the conductance has a minimum, while it reaches the maximum value when the flux is half-integer $\Phi = n+1/2$. Indeed, it costs a maximum amount of  energy to add/remove electrons to/from the dot when $\Phi = n$ and so the transport is almost blockaded and the conductance is small. On the other hand, it is energetically easy to transport electrons through the system at $\Phi = n+1/2$,  and for this reason the conductance has a maximum. Good agreement is found between the conductance calculated with the two different approaches under the condition in \cref{eq:condition}. To quantify the agreement between the two approaches, we define the error $\delta$ as 
\begin{equation}\label{eq:error delta}
\delta = \frac{\lvert G_{\rm SF} - G_{\rm ME}\rvert}{\max(G_{\rm SF}) - \min(G_{\rm SF})}\ ,
\end{equation}
and $G_{\rm SF/ME}$ denotes the conductance calculated with the SF/ME. 
The error is plotted as a function of the flux for different values of the QPC's transparency $\lvert t_{\rm QPC}\rvert^2$ at fixed temperature $\beta E_C = 10$ in \cref{fig:error filling 1}. We see that the error approaches  zero when the value of $\lvert t_{\rm QPC}\rvert^2$ decreases, such that the interferometer is more closed. However, the error does not converge to zero if the temperature is increased while keeping fixed the value of the QPC's transmission $\lvert t_{\rm QPC}\rvert^2$. 
Therefore, we can conclude that at least for describing the flux dependence of the conductance one can use the  ME to study the transport properties of an FPI in the coherent tunneling regime if i) the interferometer is sufficiently closed, and ii) such that \cref{eq:condition} is satisfied. Specifically, for a QPC transparency $\lvert t_{\rm QPC} \rvert^2 = 0.05$, the maximum error in the conductance is less than 5\%. 
, 
\\
The next quantity we compute is non-equilibrium noise, as defined in \cref{eq:noise}. We plot in \cref{fig:Noise-Fano filling 1} both   the excess noise $S-S_{eq}$ (normalized by the reference noise $S_0 \equiv 2 e^2 E_C/h$) and the  Fano factor $F$ as a function of the magnetic flux. The two quantities are calculated with both the ME (solid line, \cref{eq:noise ME matrix}) and the SF (dashed line, \cref{eq:noise SF}). We want to remark that the Fano factor obtained from the ME is independent of $\gamma$, because the noise and the current are both linear in this parameter. We see immediately that the ME and the SF give very similar results for the noise and the Fano factor, when $\lvert t_{\rm QPC}\rvert^2 \ll 1$ and when temperature and  QPC transparency satisfy \cref{eq:condition}.

In the experimental work of Ref. \cite{choi2015robust}, in addition to measuring shot noise the authors extracted an  effective charge $e^*$ from  shot-noise measurements. From the transmitted current $I$ and the noise $S$, the interfering charge $e^*$ was obtained via the formula \cite{choi2015robust,heiblum2006quantum,PhysRevLett.85.3918}
\begin{align}
S-S_{eq} = 2 e^* &I_{\rm imp} \mathcal{T}_{\rm exp} \left(1-\frac{e}{e^*}\mathcal{T}_{\rm exp}\right) \nonumber\\
&\times\left[\coth\left(\frac{\beta e^*V}{2}\right)-\frac{2}{\beta e^* V}\right], \label{eq:extract charge}
\end{align}
where the impinging current is given by $I_{\rm imp} = (e^2/h) V$, and the experimental voltage-dependent transmission probability $\mathcal{T}_{\rm exp}$ is defined as the ratio between the transmitted current and the impinging one, $\mathcal{T}_{\rm exp} = I/I_{\rm imp}$. 
By fitting \cref{eq:extract charge}  to the experimental data, an effective charge $e^*=e$ was found at filling factor $\nu = 1$, while the charge of an electron pair $e^*=2e$ was discovered at $\nu=2$ \cite{choi2015robust}.\\

\begin{figure}
	\centering
	\includegraphics[scale=0.37]{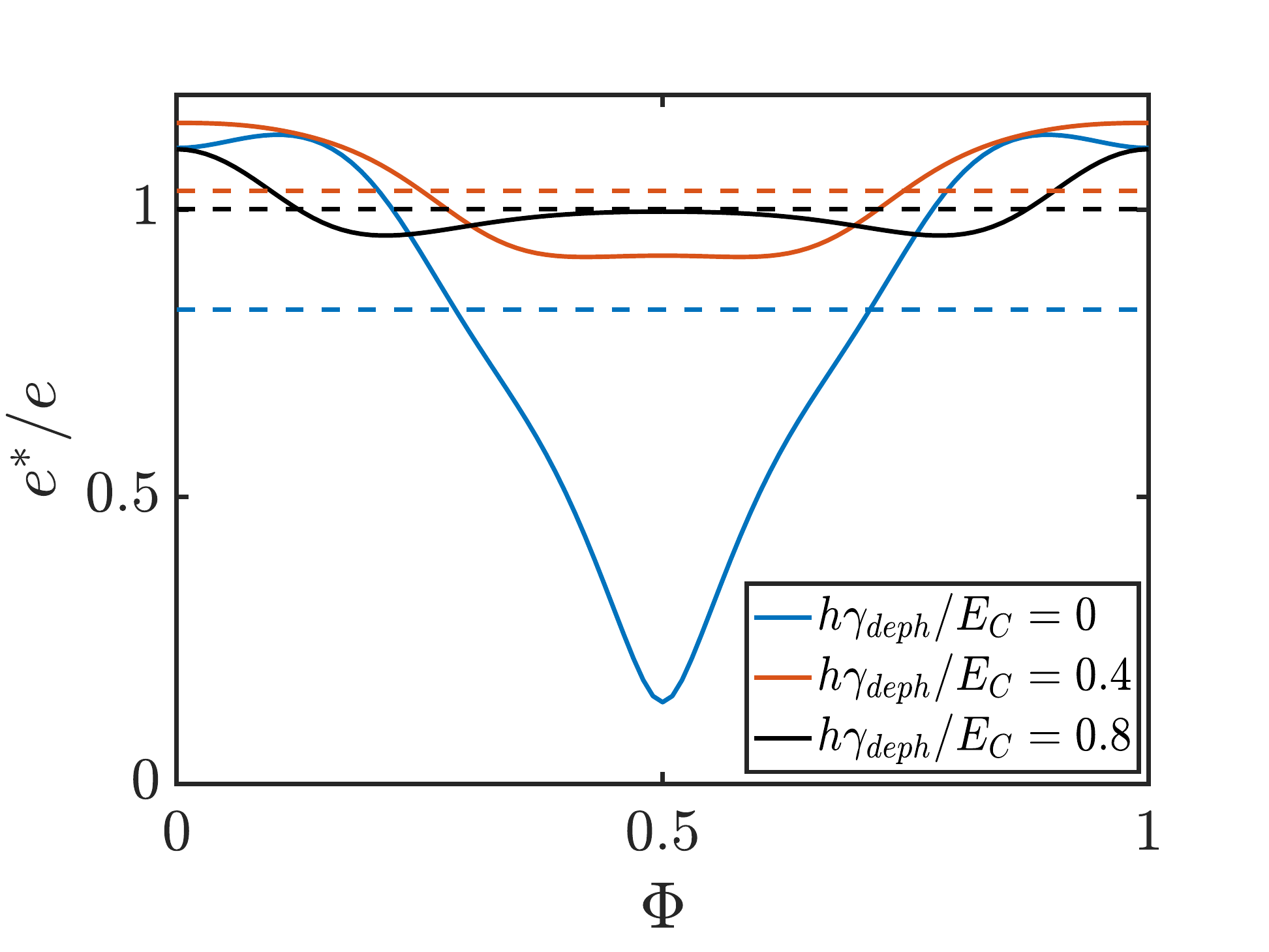}
	\caption{Effective charge obtained from \cref{eq:extract charge} as a function of the magnetic flux for a FPI at $\nu = 1$ in the fully coherent case ($\gamma_{deph.}=0$) and in presence of dephasing ($h\gamma_{deph.}/E_C=0.4, 0.8$). The horizontal lines indicate the average value of the effective charge. The average effective charge is equal to one electron, $\langle e^* \rangle \approx e$, only if the dephasing is included. Parameters are $\beta E_C = 20$, $eV/E_C=0.5$ and $\lvert t_{QPC} \rvert^2 = 0.05$.}\label{fig:effective charge main}
\end{figure}

In order to understand whether the Fano factors computed in the previous sections can be related to an effective charge, we apply the same analysis as in Ref. \cite{choi2015robust} to the predictions for voltage dependent current and excess noise computed from 
\cref{eq:current SF,eq:noise SF}, treating them as "synthetic" experimental data.
In \cref{fig:effective charge main}, we display the effective charge $e^*$ as a function of the flux for an FPI at $\nu=1$ (curve $h\gamma_{deph.}/E_C=0$), and we indicate the flux-averaged effective charge $\langle e^* \rangle$ with the dashed line. We see that in the closed limit considered here the effective charge significantly varies as a function of flux, and that its flux-averaged value is 
less than one. Clearly, the tunneling object is always one electron, and the effective charge $e^* < 1$ is a manifestation of 
interaction effects combined with the Pauli principle. 

\begin{figure}
	\centering
	\includegraphics[scale=0.37]{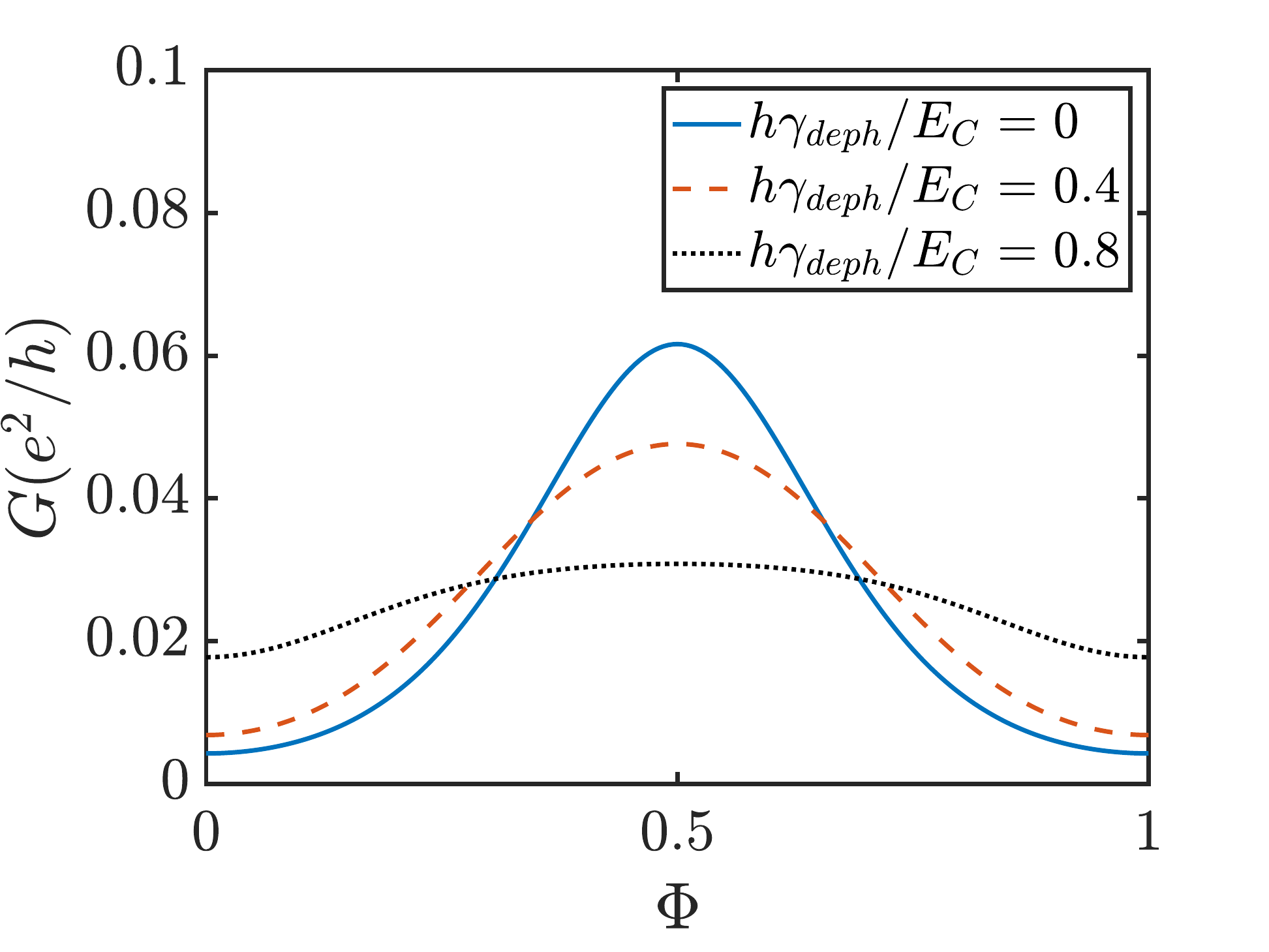}
	\caption{Conductance as a function of the magnetic flux for a FPI at $\nu = 1$ without ($\gamma_{deph.}=0$) and with ($h\gamma_{deph.}/E_C=0.4, 0.8$) dephasing for $\beta E_C = 10$, $\lvert t_{QPC} \rvert^2 = 0.05$.  The fully dephased interferometer would correspond to $h\gamma_{deph.} = E_C$.}\label{fig:conductance dephasing}
\end{figure}

\begin{figure}
	\centering
	\includegraphics[scale=0.37]{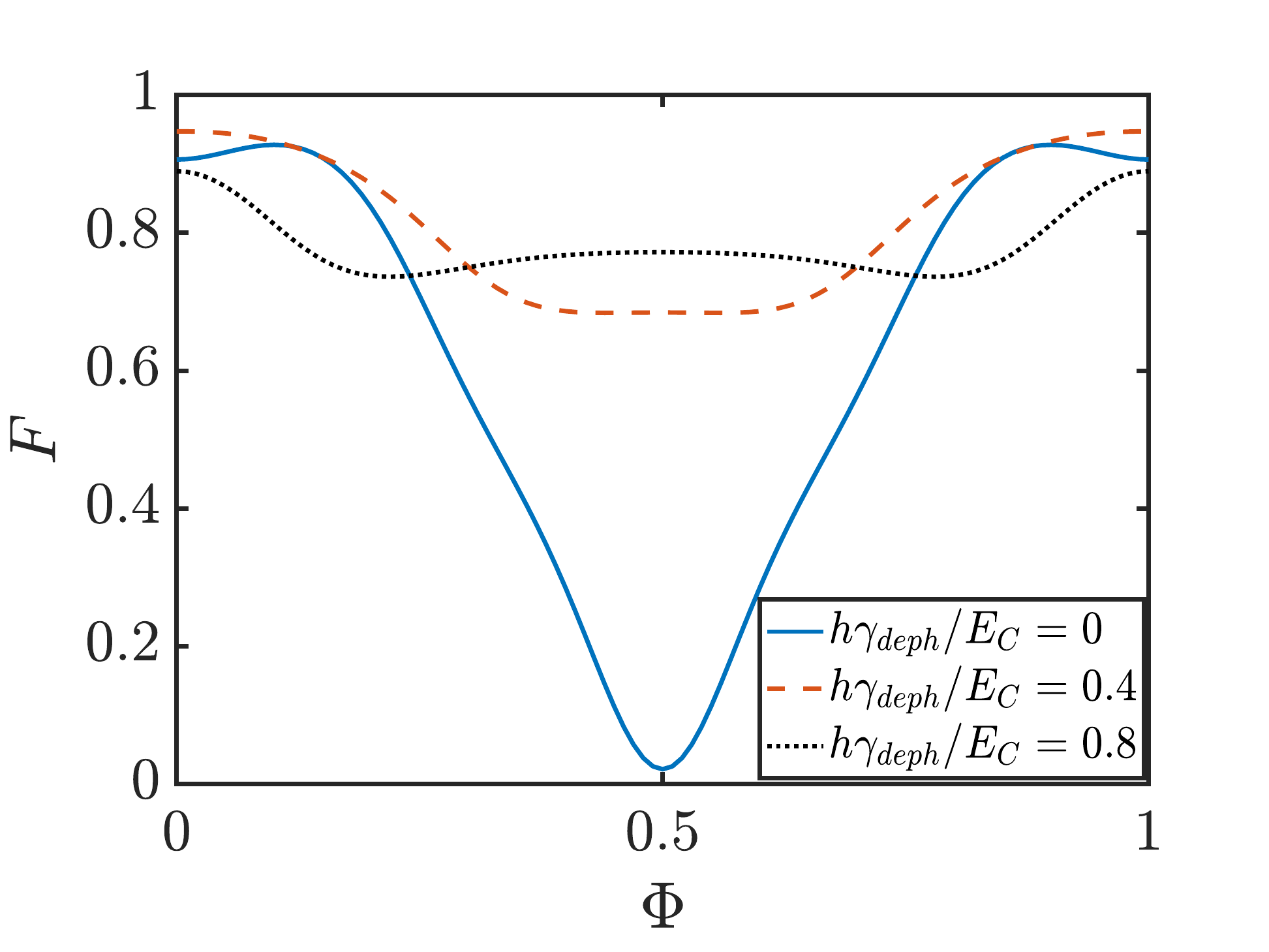}
	\caption{Fano factor as a function of the magnetic flux for a FPI at $\nu = 1$ without ($\gamma_{deph.}=0$) and with ($h\gamma_{deph.}/E_C=0.4, 0.8$) dephasing. The parameters are $\beta E_C = 20$, $eV/E_C = 0.5$ and $\lvert t \rvert^2 = 0.05$ for all the plots.}\label{fig:Fano dephasing}
\end{figure}

In experimental systems the electrons traveling along the interferometer do not completely preserve phase coherence, but are subject to dephasing. To model such  dephasing, we assume that the interfering electrons are subject to a slowly varying potential that causes the energy of the electron to fluctuate
\begin{equation}\label{eq:energy deph}
E_{deph} = E + \delta E,
\end{equation}
where $E$ is the fixed part, while $\delta E$ represents the fluctuating part of the energy. In the following, we parametrize the fluctuations as
\begin{equation}\label{eq:delta E}
\delta E = h \gamma_{deph} \varphi,
\end{equation}
where $\varphi \in [-0.5, 0.5]$ is a random variable and $\gamma_{deph}$ provides the amount of dephasing. If $\gamma_{deph} = 0$ the interferometer is fully coherent, while the totally dephased limit is obtained for $h \gamma_{deph} = E_C$. 
The effect of dephasing is to reduce the maximum of the transmission probability, and increase the width of transmission resonances. 
In \cref{fig:conductance dephasing} we show the conductance as a function of the magnetic flux in the presence of dephasing. 
The amplitude of the oscillations of the conductance is smaller for stronger dephasing, resulting in a reduction of the interference visibility.
We next generate "synthetic" experimental data by averaging the transmission probability \cref{eq:transmission probability FPI}
over energy according to \cref{eq:delta E}. We then compute  noise and Fano factor (see \cref{sec:Effective charge and dephasing} for details). We display the effect of the dephasing on the Fano factor in \cref{fig:Fano dephasing}. It is evident that the dephasing reduces the oscillations amplitude of the Fano factor. In addition, we note that the Fano factor is always smaller than one for the chosen parameters. 
Finally, we compute an effective charge by fitting our synthetic data with \cref{eq:extract charge}.  The value of the effective charge is plotted as a function of the magnetic flux in \cref{fig:effective charge main} for different values of $\gamma_{deph.}$. We indicate the respective average values with the dashed lines. We find that the inclusion of dephasing reduces the flux dependence of the effective charge, making it more meaningful to take the flux average of the effective charge.  By averaging over the flux, we obtain $\langle e^* \rangle \approx e$  when the dephasing is included. We conclude that  incorporating a dephasing mechanism into the description of a relatively closed FPI is crucial to obtain a meaningful effective charge. 
Since the interplay of interactions and dephasing in the framework of the Master equation formalism is beyond the scope of the present work, we do not present the results of our shot noise calculations in terms of an effective charge $e^*$. Instead, we rather focus on the relative enhancement of the Fano factor between the case $\nu=2$ and $\nu=1$, due to excitation of neutralons.

\section{Conclusions}
In this work we have studied an integer quantum Hall interferometer with inter-edge coupling in the strong backscattering limit. By  means of a master equation analysis, we have computed  both conductance  and noise of the system. We have found that in the presence of a strong repulsive inter-edge interaction,  neutral plasmons contribute to the electronic-transport, leading to a noticeable enhancement of excess noise in a non-equilibrium situation. In the limit of low temperature and strong inter-edge coupling, 
we have found a doubling of the Fano factor relative to the non-interacting case, indicative of correlated transmission of electrons through the interferometer. We have interpreted this enhancement of noise in terms of an electron attraction mechanism mediated by  neutral plasmons, and have argued that our results for interferometers in the strong backscattering limit are related to an enhancement of shot noise observed experimentally in more open devices.

\begin{acknowledgments}
We would like to thank B.I. Halperin for a helpful discussion. This work was supported by the IMPRS MiS Leipzig and by DFG grant RO 2247/11-1. 
\end{acknowledgments}

\appendix
\onecolumngrid
\section{Calculation of matrix element \label{app:suppl calculation of matrix element}}
We provide here some more details about the calculation of the matrix element for the tunneling of one electron in a FPI. The fermionic operator $\psi_{1,D}(x)$, responsible for the annihilation of one electron in the interfering edge of the dot at position $x$, has the form \cite{Geller1997}
\begin{equation}
\psi_{1,D}(x) = \frac{1}{\sqrt{2\pi \alpha}}e^{-i\chi_{1}} e^{2\pi i N_{1} x/L} e^{i\phi^p_{1,D}(x)} = \frac{1}{\sqrt{2\pi \alpha}}e^{-i\chi_{1}} e^{2\pi i N_{1} x/L} e^{i\left(\phi^p_\rho(x)-\phi^p_\sigma(x)\right)/\sqrt{2}}.
\end{equation}
The needed matrix element is
\begin{align}
&\langle N_1, N_2, \{n\}_\rho, \{n\}_\sigma | \psi_{1,D}(x) |N+1, N_2, \{n^\prime\}_\rho, \{n^\prime\}_\sigma \rangle = \nonumber\\
&= \frac{1}{\sqrt{2\pi\alpha}} \langle N_1| e^{-i\chi_1} e^{2\pi i N_1 x/L} |N_1+1\rangle \langle\{n\}_\rho | e^{i\phi^p_\rho(x)/\sqrt{2}} |\{n^\prime\}_\rho\rangle \langle\{n\}_\sigma | e^{-i\phi^p_\sigma(x)/\sqrt{2}} |\{n^\prime\}_\sigma\rangle \nonumber\\
&= \frac{c}{\sqrt{2\pi\alpha}} e^{2\pi i (N_1+1) x/L} \langle\{n\}_\rho | e^{i\phi^p_\rho(x)/\sqrt{2}} |\{n^\prime\}_\rho\rangle \langle\{n\}_\sigma | e^{-i\phi^p_\sigma(x)/\sqrt{2}} |\{n^\prime\}_\sigma\rangle, \label{s-eq:matrix element filling2}
\end{align}
where $c$ is a phase factor, $|c|^2 = 1$ and we used that $e^{-i\chi}|N+1\rangle = c|N\rangle$, that follows from the commutation relation between $\chi$ and $N$, $[\chi, N] = i$. 
The matrix element for plasmon excitations is
\begin{equation}\label{s-eq:plasmon matrix element}
	\langle\{n\} | e^{i\phi^p(x)/\sqrt{2}} |\{n^\prime\}\rangle = \langle\{n\} | e^{i \sum_{k>0} \left(\beta_k b_k + \beta_k^* b_k^\dagger\right)} |\{n^\prime\}\rangle
	= \prod_{k>0} \langle n_k | e^{i  \left(\beta_k b_k + \beta_k^* b_k^\dagger\right)} |n_k^\prime\rangle,
\end{equation}
where $b_k,b_k^\dagger$ are bosonic operators satisfying $[b_k, b_q] = 0$, $[b_k, b^\dagger_q] = \delta_{k,q}$, and we defined
\begin{equation}\label{s-eq:plasmon field}
	\beta_k = \sqrt{\frac{\pi}{kL}}e^{ikx-\alpha k/2}.
\end{equation}
Now, we have for $n_k > n_k^\prime$
\begin{align}\label{s-eq:one plasmon}
\langle n_k | e^{i  \left(\beta_k b_k + \beta_k^* b_k^\dagger\right)} |n_k^\prime\rangle &= e^{|\beta_k|^2/2} \langle n_k | e^{i  \beta_k b_k} e^{i \beta_k^* b_k^\dagger} |n_k^\prime\rangle \nonumber\\
&= e^{|\beta_k|^2/2}  \sum_{l,m}\frac{\left(i\beta_k\right)^l}{l!} \frac{\left(i\beta_k^*\right)^m}{m!} \langle n_k | b_k^l \left(b_k^\dagger\right)^m|n_k^\prime\rangle \nonumber\\
&= e^{|\beta_k|^2/2} \sum_{l,m}\frac{\left(i\beta_k\right)^l}{l!} \frac{\left(i\beta_k^*\right)^m}{m!} \frac{1}{\sqrt{n_k! n_k^\prime !}}\langle 0 | b_k^{(l+n_k)} \left(b_k^\dagger\right)^{(m + n_k^\prime)} |0\rangle \nonumber\\
&= e^{|\beta_k|^2/2} \sum_{l,m}\frac{\left(i\beta_k\right)^l}{l!} \frac{\left(i\beta_k^*\right)^m}{m!} \sqrt{\frac{(l+n_k)! (m+n_k^\prime)!}{n_k! n_k^\prime !}}\langle l+n_k |m+n_k^\prime \rangle \nonumber\\
&= e^{|\beta_k|^2/2} \sum_{l}\frac{\left(i\beta_k\right)^l}{l!} \frac{\left(i\beta_k^*\right)^{l+n_k-n_k^\prime}}{(l+n_k-n_k^\prime)!} \frac{(l+n_k)!}{\sqrt{n_k! n_k^\prime !}} \nonumber\\
&= e^{|\beta_k|^2/2} \frac{\left(i\beta_k^*\right)^{n_k-n_k^\prime}}{\sqrt{n_k! n_k^\prime !}} \sum_l \frac{(l+n_k)!}{l! (l+n_k-n_k^\prime)!} \left(-\left|\beta_k\right|^{2}\right)^l \nonumber\\
&= e^{|\beta_k|^2/2} \frac{\left(i\beta_k^*\right)^{n_k-n_k^\prime}}{(n_k-n_k^\prime) !} \sqrt{\frac{n_k!}{n_k^\prime!}}~\Phi\left(n_k+1, n_k-n_k^\prime+1, -\left|\beta_k\right|^2\right) \nonumber\\
&= e^{-|\beta_k|^2/2} \left(i\beta_k^*\right)^{n_k-n_k^\prime} \sqrt{\frac{n_k^\prime!}{n_k!}}~L_{n_k^\prime}^{n_k-n_k^\prime}\left(\left|\beta_k\right|^2\right),
\end{align}
where we used the Baker-Campbell-Hausdorff formula, the power series of the exponential, the orthonormal states
\begin{equation}
|n_k\rangle = \frac{\left(b_k^\dagger\right)^{n_k}}{\sqrt{n_k!}} |0\rangle,
\end{equation}
the definition of the confluent hypergeometric function $\Phi$ and its relation to the associated Laguerre polynomials $L_a^b(x)$
\begin{align}
\Phi\left(n_k+1, n_k-n_k^\prime+1, -\left|\beta_k\right|^2\right) &=  \frac{(n_k-n_k^\prime)!}{n_k!} \sum_l \frac{(l+n_k)!}{l! (l+n_k-n_k^\prime)!} \left(-\left|\beta_k\right|^{2}\right)^l =\frac{n_k^\prime! (n-n_k^\prime)!}{n_k!} e^{-|\beta_k|^2} L_{n_k^\prime}^{n_k-n_k^\prime}\left(\left|\beta_k\right|^2\right).
\end{align}
By inserting \cref{s-eq:one plasmon} in \cref{s-eq:plasmon matrix element}, taking the modulus square, using \cref{s-eq:plasmon field} and the momentum quantization $k = 2\pi m/L$, we get
\begin{align}\label{s-eq:plasmon matrix element 2}
\left|\langle\{n\} | e^{i\phi^p(x)/\sqrt{2}} |\{n^\prime\}\rangle\right|^2 &= 
\prod_{k>0} e^{-\lvert \beta_k \rvert^2} \lvert \beta_k \rvert^{2|n_k-n_k^\prime|} \left(\frac{n^{(<)}_k!}{n^{(>)}_k!}\right)
\left[L_{n^{(<)}_k}^{|n_k-n_k^\prime|} \left(\lvert \beta_k \rvert^2\right)\right]^2 \nonumber\\
&=	\prod_{m=1}^{+\infty} e^{-(1/2m) e^{-2\pi\alpha m/L}} \left(\frac{1}{2m}  e^{-2\pi\alpha m/L}\right)^{|n_m-n_m^\prime|} \left(\frac{n^{(<)}_m!}{n^{(>)}_m!}\right) 
\left[L_{n^{(<)}_m}^{|n_m-n_m^\prime|} \left(\frac{1}{2m}  e^{-2\pi\alpha m/L}\right)\right]^2.
\end{align}
Let suppose that $n_m = n_m^\prime = 0$ for every $m>m_{\rm max}$. Then, we can write \cref{s-eq:plasmon matrix element 2} as
\begin{align}
&	\lvert\langle\{n\} | e^{i\phi^p(x)/\sqrt{2}} |\{n^\prime\}\rangle\rvert^2 = \nonumber\\
&\prod_{m=1}^{m_{\rm max}} e^{-(1/2m) e^{-2\pi\alpha m/L}} \left(\frac{1}{2m}  e^{-2\pi\alpha m/L}\right)^{|n_{m}-n_{m}^\prime|} \frac{n^{(<)}_{m}!}{n^{(>)}_{m}!} 
\left[L_{n^{(<)}_{m}}^{|n_{m}-n_{m}^\prime|} \left(\frac{1}{2m}  e^{-2\pi\alpha m/L}\right)\right]^2 
\prod_{m>m_{\rm max}} e^{-(1/2m) e^{-2\pi\alpha m/L}} \nonumber\\
&= \prod_{m=1}^{m_{\rm max}} e^{-(1/2m) e^{-2\pi\alpha m/L}} \left(\frac{1}{2m}  e^{-2\pi\alpha m/L}\right)^{|n_{m}-n_{m}^\prime|} \frac{n^{(<)}_{m}!}{n^{(>)}_{m}!} 
\left[L_{n^{(<)}_{m}}^{|n_{m}-n_{m}^\prime|} \left(\frac{1}{2m}  e^{-2\pi\alpha m/L}\right)\right]^2  \frac{\prod_{m=1}^{+\infty} e^{-(1/2m) e^{-2\pi\alpha m/L}}}{\prod_{m=1}^{m_{\rm max}} e^{-(1/2m) e^{-2\pi\alpha m/L}}} \nonumber\\
&= \sqrt{1-e^{-2\pi\alpha/L}} 
\frac{\prod_{m=1}^{m_{\rm max}} e^{-(1/2m) e^{-2\pi\alpha m/L}} \left(\frac{1}{2m}  e^{-2\pi\alpha/L}\right)^{|n_{m}-n_{m}^\prime|} \frac{n^{(<)}_{m}!}{n^{(>)}_{m}!} 
	\left[L_{n^{(<)}_{m}}^{|n_{m}-n_{m}^\prime|} \left(\frac{1}{2m}  e^{-2\pi\alpha m/L}\right)\right]^2}{\prod_{m=1}^{m_{\rm max}} e^{-(1/2m) e^{-2\pi\alpha m/L}}},\label{s-eq:final matrix}
\end{align}
and we obtain the final result by inserting \cref{s-eq:final matrix} into \cref{s-eq:matrix element filling2} and taking the limit $\alpha \to 0$.

\twocolumngrid
\section{Effective charge and dephasing \label{sec:Effective charge and dephasing}}
\begin{figure}
	\centering
	\includegraphics[scale=0.37]{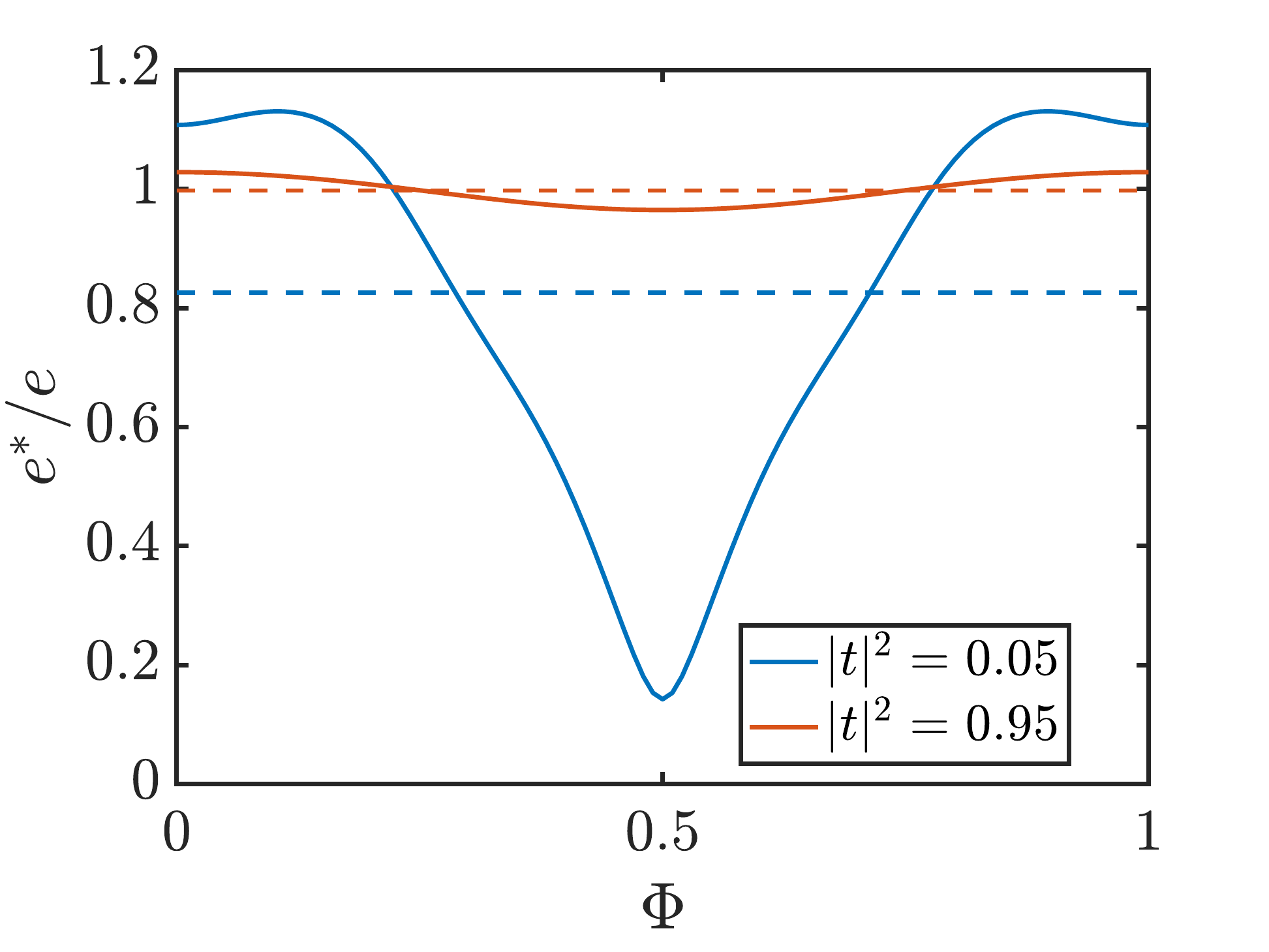}
	\caption{Effective charge as a function of the magnetic flux obtained from \cref{eq:extract charge} in the closed limit ($\lvert t \rvert^2 = 0.05$) and in the open limit ($\lvert t \rvert^2 = 0.95$) at voltage $eV/E_C = 0.5$ and temperature $\beta E_C = 20$. The dashed lines represent the flux-averaged effective charge.}\label{fig:charge filling 1}
\end{figure}

In this appendix, we provide some more details about the inclusion of the dephasing in extracting an effective charge from excess noise.
First, we show in \cref{fig:charge filling 1} a comparison between the values for  effective charges at $\nu=1$  in the closed and in the open limit. The flux-averaged effective charge is indicated with the dashed line. In the closed limit, the effective charge strongly oscillates with the flux. On the other hand, the oscillations of $e^*$ are much less pronounced in the open limit, and we find the expected effective charge $e^* \approx e$. Therefore, the standard procedure in \cref{eq:extract charge} to extract the effective charge from the shot noise works well in the open limit but
does not give useful results when  the interferometer operates in the closed limit. For this reason we do not convert our predictions for excess noise at $\nu=2$ into effective charges. 

\begin{figure}
	\centering
	\includegraphics[scale=0.37]{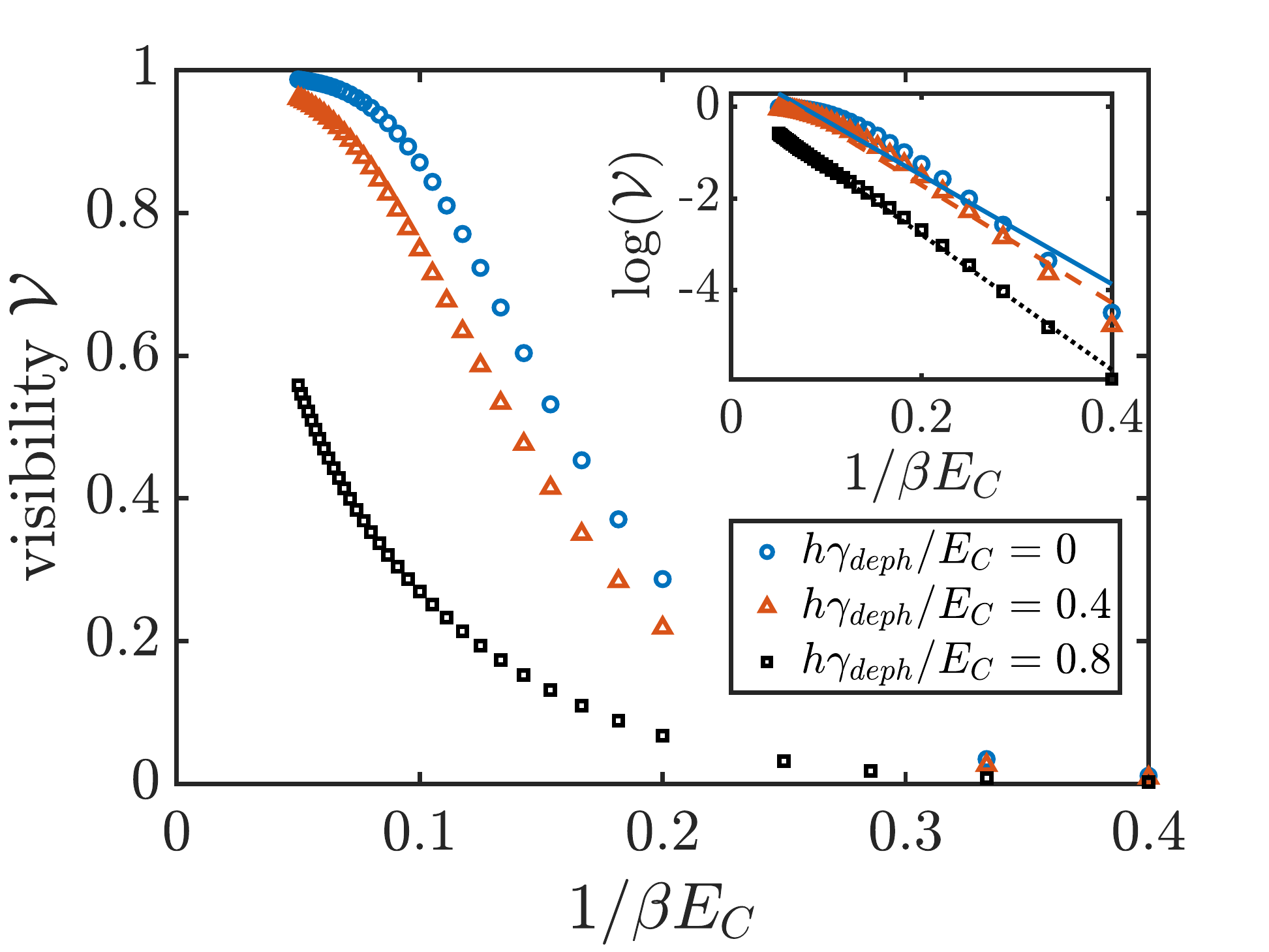}
	\caption{Visibility, defined in \cref{eq:visibility}, of a FPI at $\nu=1$ as a function of the temperature for different values of $\gamma_{deph.}$. In the inset the logarithm of the visibility is plotted as a function of the temperature.}\label{fig:visibility}
\end{figure}

When including dephasing into the description of the FPI, an effective charge can be extracted for interferometers operating in the closed limit as well. According to \cref{eq:delta E}, the transmission probability for a FPI subject to the dephasing is
\begin{equation}\label{eq:T varphi}
\mathcal{T}(E, \gamma_{\rm deph.}, \varphi) = \frac{\lvert t \rvert^4}{\left\lvert 1+r^2 e^{2\pi i \left(\frac{E}{E_C} + \Phi + \frac{h\gamma_{deph.}}{E_C}\varphi\right)}\right\rvert^2},
\end{equation}
and by averaging over the random variable $\varphi$ we have
\begin{equation}\label{eq:T deph}
\mathcal{T}(E, \gamma_{deph.}) = \int_{-0.5}^{0.5} d\varphi~\mathcal{T}(E, \gamma_{deph.}, \varphi).
\end{equation}
The conductance is obtained from \cref{eq:T deph} as
\begin{equation}
G = \frac{e^2}{h}\int dE~\mathcal{T}(E, \gamma_{deph.}) \left(-\frac{\partial f}{\partial E}\right).
\end{equation}
To quantify the quality of the interference signal, we define the visibility $\mathcal{V}$ as
\begin{equation}\label{eq:visibility}
	\mathcal{V} = \frac{\max\left(G\right) - \min\left(G\right)}{\max\left(G\right) + \min\left(G\right)}.
\end{equation}
We plot the visibility as a function of temperature for different values of $\gamma_{deph.}$ in the right panel of \cref{fig:visibility}. We can immediately see that the presence of dephasing strongly reduces the visibility. In the absence of dephasing, the visibility decays exponentially with temperature (see \cref{fig:visibility}), while in the presence of dephasing deviations from an exponential decay are visible. Overall, the  visibility decays faster when the dephasing becomes stronger.

\begin{figure}
	\centering
	\includegraphics[scale=0.37]{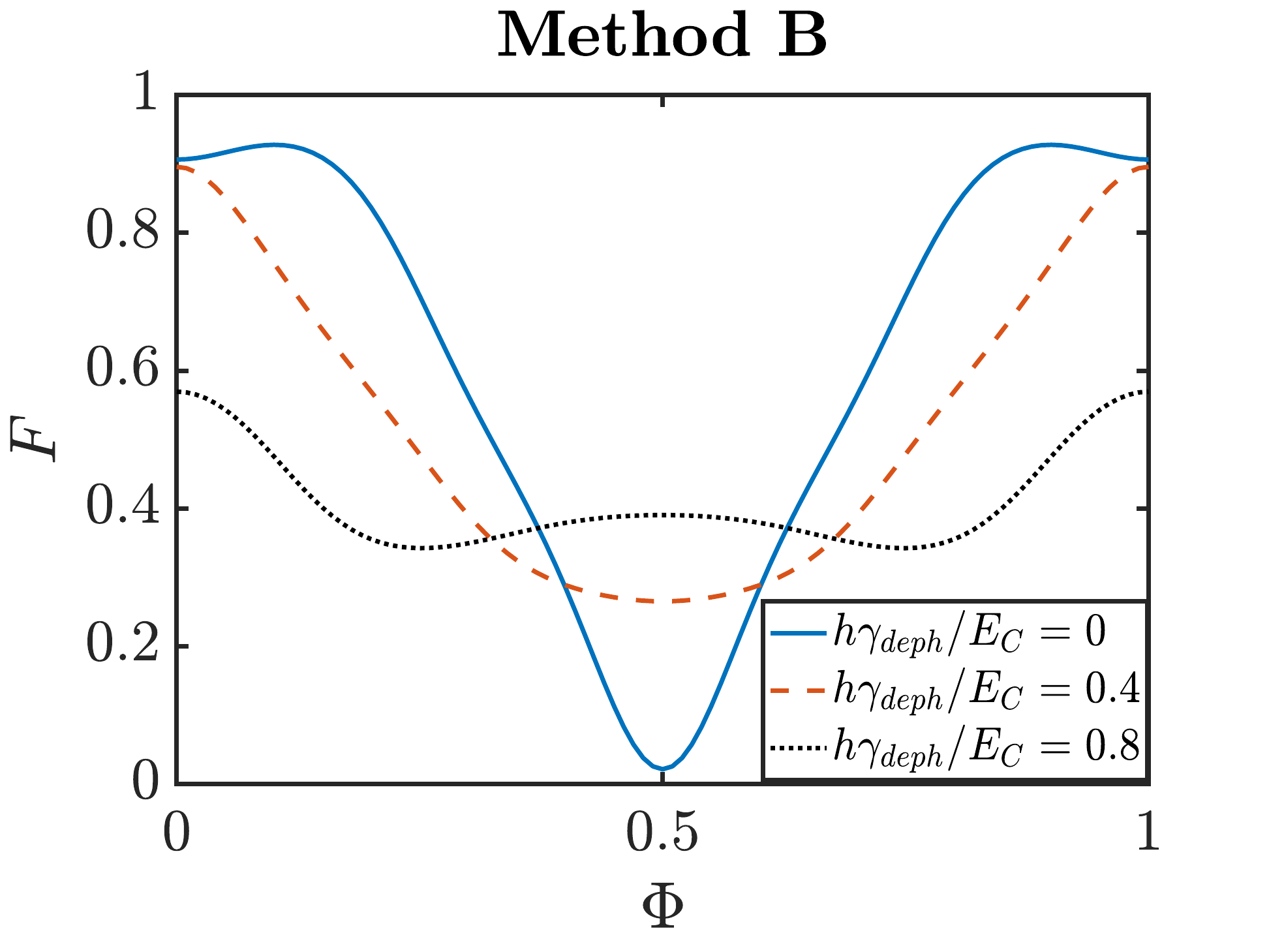}
	\caption{Fano factor as a function of the magnetic flux in the presence of dephasing ($h\gamma_{deph.}/E_C=0.4, 0.8$), calculated from \cref{eq:noise dephased 2}. The parameters are $\beta E_C = 20$, $eV/E_C = 0.5$ and $\lvert t \rvert^2 = 0.05$.}\label{fig:noise dephasing}
\end{figure}

We now want to calculate the excess noise in the presence of the dephasing modeled in \cref{eq:energy deph,eq:delta E}. In the main text, we first averaged the transmission probability $\mathcal{T}(E, \gamma_{deph.},\varphi)$ over $\varphi$, and then calculate the noise from it. Therefore, the dephased noise is similar to \cref{eq:noise SF} but with the transmission probability given by \cref{eq:T deph}
\begin{align}
S = \frac{2e^2}{h} &\int dE~\{ \mathcal{T}(E, \gamma_{deph.}) \left[f_L(1-f_L) + f_R (1-f_R)\right] \nonumber\\
&+ \mathcal{T}(E, \gamma_{deph.}) \left[1-\mathcal{T}(E, \gamma_{deph.})\right](f_L-f_R)^2\}.\label{eq:noise dephased 1}
\end{align}
The Fano factor obtained from \cref{eq:noise dephased 1} is shown in \cref{fig:Fano dephasing} and the effect of the dephasing is to diminish the amplitude of its oscillations.\\
We now discuss a different method (method B) to calculate the noise in presence of dephasing. This consists in calculating the noise for a given $\varphi$ and just at the end the average over $\varphi$ is performed. Accordingly, the noise is given by
\begin{widetext}
\begin{equation}
S = \frac{2e^2}{h} \int d\varphi \int dE~\{ \mathcal{T}(E, \gamma_{deph.}, \varphi) \left[f_L(1-f_L) + f_R (1-f_R)\right]
+ \mathcal{T}(E, \gamma_{deph.}, \varphi) \left[1-\mathcal{T}(E, \gamma_{deph.}, \varphi)\right](f_L-f_R)^2\}, \label{eq:noise dephased 2}
\end{equation}
\end{widetext}
with $\mathcal{T}(E, \gamma_{deph.}, \varphi)$ given by \cref{eq:T varphi}. \Cref{eq:noise dephased 2} corresponds to averaging the non-dephased noise in \cref{eq:noise SF} over the flux $\Phi$ in a window of length $h\gamma_{deph.}/E_C$, as it can be seen from \cref{eq:T varphi} and the change of variable $\Phi^\prime = \Phi + (h\gamma_{deph.}/E_C)\varphi$. This way of including the dephasing would be directly applicable also to the master equation.
Accordingly, the Fano factor obtained from \cref{eq:noise dephased 2} is shown in \cref{fig:noise dephasing}. As an effect of the dephasing, the amplitude of the oscillations of $F$ gets smaller.

\begin{figure}
	\centering
	\includegraphics[scale=0.37]{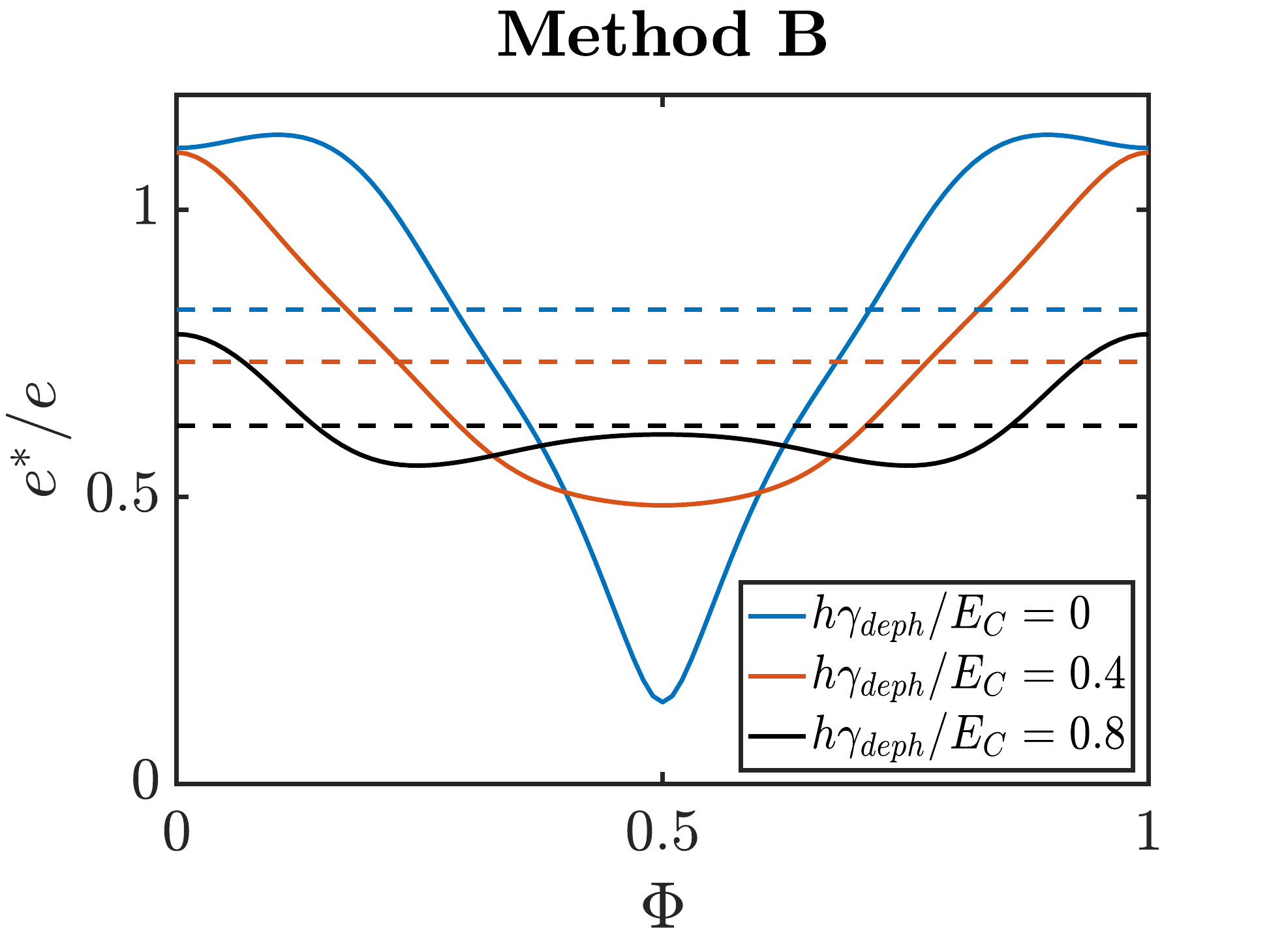}
	\caption{Effective charge extracted from the noise calculated from \cref{eq:noise dephased 2}. The horizontal lines indicate the average value of the effective charge. The effective charge is reduced when the dephasing is stronger. Parameters are $\beta E_C = 20$, $eV/E_C=0.5$ and $\lvert t_{QPC} \rvert^2 = 0.05$.}\label{fig:effective charge}
\end{figure}

Finally, we compute the effective charge by using \cref{eq:extract charge}, as in Ref. \cite{choi2015robust}, from the noise calculated with the two different methods. In the main text (see \cref{fig:effective charge main}), we have shown the effective charge obtained from the noise calculated with \cref{eq:noise dephased 1}, and we have seen that the dephasing helps to get $\langle e^* \rangle =e$.
The effective charge is plotted as a function of the magnetic flux in \cref{fig:effective charge} for method B and different values of $\gamma_{deph.}$. We indicate the respective average values with the dashed lines. We find that in contrast to the results in \cref{fig:effective charge main}, 
the average effective charge $\langle e^* \rangle$ gets smaller for stronger dephasing, when \cref{eq:noise dephased 2} is used to obtain the noise. Therefore, \cref{eq:extract charge} and \cref{eq:noise dephased 2} are not compatible with each other. Since method B is not helpful in extracting an effective charge with a value close to the electron charge in the test case $\nu=1$, we do not attempt to compute an effective charge in the more complicated case $\nu=2$, and rather present the relative enhancement of the Fano factor as our main result. 
%

\bibliography{biblio_v6}

\end{document}